\documentclass[preprint,showpacs,12pt,floatfix,aps]{revtex4}
\usepackage{graphicx}
\usepackage{latexsym}
\usepackage{amsmath}
\input psfig.sty

\newcommand{\bq}{\begin{equation}}
\newcommand{\eq}{\end{equation}}
\newcommand{\bqn}{\begin{eqnarray}}
\newcommand{\eqn}{\end{eqnarray}}
\newcommand{\nb}{\nonumber}
\newcommand{\lb}{\label}
\newcommand{\rr}{\bf r}

\begin{document}
\title{Gravastars with an Interior Dark Energy Fluid Forming a Naked Singularity}
\author{C. F. C. Brandt $^{2}$}
\email{fredcharret@yahoo.com.br}
\author{R. Chan $^{1}$}
\email{chan@on.br}
\author{M.F.A. da Silva $^{2}$}
\email{mfasnic@gmail.com}
\author{P. Rocha $^{2,3}$}
\email{pedrosennarocha@gmail.com}
\affiliation{\small $^{1}$ Coordena\c{c}\~ao de Astronomia e Astrof\'{\i}sica, 
Observat\'orio Nacional, Rua General Jos\'e Cristino, 77, S\~ao Crist\'ov\~ao  
20921-400, Rio de Janeiro, RJ, Brazil\\
$^{2}$ Departamento de F\'{\i}sica Te\'orica, Instituto de F\'{\i}sica, 
Universidade do Estado do Rio de Janeiro, Rua S\~ao Francisco Xavier 524, 
Maracan\~a 20550-900, Rio de Janeiro, RJ, Brasil\\
$^3$ IST - Instituto Superior de Tecnologia de Paracambi, FAETEC, Rua Sebasti\~ao de Lacerda
s/n, Bairro da F\'abrica, Paracambi, 26600-000, RJ, Brazil}
 
\date{\today}

\begin{abstract}
We consider a gravastar model made of anisotropic dark energy  
with an infinitely thin spherical shell of a perfect fluid with the equation of 
state $p = (1-\gamma)\sigma$ 
with an external de Sitter-Schwarzschild region. 
It is found that in some cases the  models represent the "bounded excursion"  
stable gravastars, where  the thin shell is oscillating between two finite radii, 
while in other cases they collapse until the formation of black holes
or naked singularities.  An interesting result is that we can have
black hole and stable gravastar formation even with an interior and a shell constituted of dark
and repulsive dark energy, as also shown in previous work. 
Besides, in one case we have a dynamical evolution to a black hole (for $\Lambda=0$) 
or to a naked singularity (for  $\Lambda > 0$). This is the first time in the literature that a naked
singularity emerges from a gravastar model.

\end{abstract}

\pacs{98.80.-k,04.20.Cv,04.70.Dy}

\maketitle

\section{Introduction}

Nowadays, although we know that gravastars are not an alternative to black holes, 
these theoretical objects cannot be ignored and, in fact, they have received a 
considerable attention \cite{grava}\cite{JCAP}-\cite{JCAP4}.

The pioneer model of gravastar was proposed by Mazur and Mottola (MM) \cite{MM01},
consisting of five layers: an internal core
$0 < R < R_1$ , described by the de Sitter universe, an intermediate thin layer of stiff fluid
$R_1 < R < R_2$ , an external region $R > R_2$, described by the Schwarzschild solution, and two 
infinitely thin shells, appearing, respectively, on the hypersurfaces $R = R_1$ and
$R = R_2$, where $R$ denotes the radius of the star. The intermediate layer is constructed in such way that $R_1$ is inner than the de Sitter horizon, while $R_2$ is outer than the Schwarzschild horizon, eliminating the apparent horizon. Configurations with a de Sitter interior have long history which we can find, for example, in the work of Dymnikova and Galaktionov \cite{irina}.  
After this work, Visser and Wiltshire \cite{VW04} pointed out that there are 
two different types of stable gravastars which are stable gravastars and 
``bounded excursion" gravastars. In the spherically symmetric case, the motion 
of the surface of the gravastar can be written in the form \cite{VW04},
\bq
\lb{1.4}
\frac{1}{2}\dot{R}^{2} + V(R) = 0,
\eq
where $\dot{R} \equiv dR/d\tau$, with 
$\tau$ being the proper time of the surface. Depending on the properties of 
the potential $V(R)$, we can have two kinds of gravastars, which are (i) Stable gravastars and 
(ii)"Bounded excursion" gravastars. For the former one there must exist a radius $R_{0}$ such that $V\left(R_{0}\right) = 0$,  $V'\left(R_{0}\right) = 0$, $V''\left(R_{0}\right) > 0$, where a prime denotes the ordinary differentiation with respect to the indicated argument.
If and only if there exists such a radius $R_{0}$ for which the above conditions are satisfied,
the model is said to be stable. 

On the other hand, for the second one, there exist two radii $R_{1}$ and $R_{2}$ such that $V\left(R_{1}\right) = 0$, $V'\left(R_{1}\right) \le 0$, $V\left(R_{2}\right) = 0$, $V'\left(R_{2}\right) \ge 0$, with $V(R) < 0$ for $a \in \left(R_{1}, R_{2}\right)$, where $R_{2} > R_{1}$. 

Many authors have studied the stability of the gravastar models for several equations of state, 
among them we can  mention \cite{VW04} - \cite{JCAP1}. Some generalizations of these models 
can be found in the literature \cite{JCAP2} - \cite {Lobo}.

The study of gravastar, in general, has considered these objects embedded in a Schwarzschild spacetime 
(except in the references \cite{Carter05} and \cite{JCAP3}). However, taking the cosmological 
point of view 
that the universe must be fullfiled by a considerable amount of dark energy, it is very 
important to investigate its influence in the gravastar stability and in its possible dynamic evolution. 
In a first step, we have considered the de Sitter-Schwarzschild exterior spacetime, in order to 
introduce a positive cosmological constant, which has been suggested as a dark energy candidate.

In this paper, we generalize our previous works \cite{JCAP3}-\cite{arxiv} to the case where the 
equation of state of the infinitely thin shell is still given by $p= (1-\gamma)  \sigma$ with 
$\gamma$ being a constant, the interior consists of an anisotropic dark energy fluid 
(similarly used in \cite{arxiv}), 
while the exterior is now the de Sitter-Schwarzschild space (similarly used in \cite{JCAP3}). 
In the previous work \cite{arxiv} we showed that anisotropic dark energy could collapse 
forming a black hole or a gravastar. 
We also studied the effect of anisotropy of the interior fluid on the formation of gravastars and it 
was concluded that the sign of  the difference between the pressures (radial and tangential)  
affects the conditions of the formation of the gravastar and black holes when the interior fluid 
of prototype gravastars is anisotropic. This is confirmed in this present work.

Here we shall first construct three-layer dynamical models,  and then show both types of
gravastars and black holes exist for various situations. In addition, this model 
shows that a naked singularity can be the final stage of a gravastar. The rest of 
the paper is  organized as follows: In Sec. II we present the metrics of the 
interior and exterior spacetimes, and write down the motion of the thin shell
in the form of equation (\ref{1.4}).  In Sec. III we show the definitions of dark and 
phantom energy, for the
interior fluid and for the shell.  In Sec. IV we discuss the different 
structures that are formed from standard or dark energy anisotropic fluid. 
Finally, in Sec. V we present our conclusions.

\section{ Dynamical Three-layer Prototype Gravastars}

The interior fluid is made of an anisotropic dark energy fluid with a metric
given by the first Lobo's model \cite{Lobo}
\bq
ds^2_{i}=-f_1 dt^2 + f_2 dr^2 + r^2 d\Omega^2,
\lb{ds2-}
\eq
where  $d\Omega^2 \equiv d\theta^2 + \sin^2(\theta)d\phi^2$, and 
\bqn
f_1 &=& (1-2 a r^2)^{-\frac{1+3\omega}{2}},\nb\\
f_2 &=& \frac{1}{1 -2 a r^2},
\eqn  
where $\omega$ is a constant, and its physical meaning can be seen from the
following equation (\ref{prpt}). Since the mass is given by 
$\bar m(r)=4\pi\rho_0 r^3/3$ and $a=4\pi\rho_0/3$ then we have that $a > 0$,
where $\rho_0$ is the homogeneous energy density.
Note that there is a horizon at $r_h=1/\sqrt{2a}$, thus the radial coordinate
must obey $r < r_h$. 
The corresponding energy density $\rho$, radial  and tangential  pressures $p_r$ and 
$p_t$ are given, respectively, by
\bqn
\rho&=&\rho_0=constant, \nb \\
p_r&=&\omega \rho_0, \nb \\
p_t&=&\omega \rho_0 \left[ 1 + \frac{4\pi}{6} \frac{(1+\omega)(1+3\omega)\rho_0 r^
2}
{\omega \left(1- \frac{8\pi \rho_0}{3}r^2 \right)} \right], \\
\lb{prpt}
\end{eqnarray}
when $\omega=-1$ and $\omega=-1/3$ we obtain an interior isotropic pressure fluid.

The exterior spacetime is given by the de Sitter-Schwarzschild metric
\bq
ds^2_{e}= - f dv^2 + f^{-1} d{\rr}^2 + {\rr}^2 d\Omega^2,
\lb{ds2+}
\eq
where $f=1 - \frac{2m}{\rr}-({\rr}/L_e)^2$ and $L_e=\sqrt{3/\Lambda_e}$.
The metric of the hypersurface  on the shell is given by
\bq
ds^2_{\Sigma}= -d\tau^2 + R^2(\tau) d\Omega^2,
\lb{ds2Sigma}
\eq
where $\tau$ is the proper time.

Since $ds^2_{i} = ds^2_{e} = ds^2_{\Sigma}$, we find that  $r_{\Sigma}={\rr}_{\Sigma}=R$,
and  
\bqn
\lb{dott2}
f_1\dot t^2 - f_2 \dot r^2_{\Sigma} &=&  1,\\
\lb{dotv2}
f\dot v^2 - \frac{\dot {\rr}^2_{\Sigma}}{f} &=& 1,
\eqn
where the dot denotes the ordinary differentiation with respect to the proper time.
On the other hand,  the interior and exterior normal vectors to the thin shell are given by
\bqn
\lb{nalpha-}
n^{i}_{\alpha} &=& (-\dot r_{\Sigma}, \dot t, 0 , 0 ),\nb\\
n^{e}_{\alpha} &=& (-\dot {\rr}_{\Sigma}, \dot v, 0 , 0 ).
\eqn
Then, the interior and exterior extrinsic curvatures, using the junction condition 
$r_{\Sigma}={\rr}_{\Sigma}=R$ and equations (\ref{dott2}), (\ref{dotv2}) calculated
at the shell radius, are given by
\bqn
K^{i}_{\tau\tau} &=& -(1-2 a R^2)^{-(3\omega+1)/2} \left\{ \left[ 6 (1-2 a 
R^2)^{(3\omega+1)/2} \dot R^2 \omega+6 a R^2 \dot t^2 \omega+2 a 
R^2 \dot t^2-3 \dot t^2 \omega-\dot t^2 \right] \right. \nb \\
& &\left. \times a R \dot t- (1-2 a R^2)^ {(3 
\omega+1)/2} (-1+2 a R^2) (\dot R \ddot t-\ddot R \dot t)\right\} (-1+2 a R^2)
^{-1}, 
\lb{Ktautau-}
\eqn
\bq
K^{i}_{\theta\theta} =
\dot t(1-2a R^2) R,
\lb{Kthetatheta-}
\eq
\bq
K^{i}_{\phi\phi} = K^{i}_{\theta\theta}\sin^2(\theta),
\lb{Kphiphi-}
\eq
\bqn
K^{e}_{\tau\tau}&=&\dot v [(2 L_e^2 m \dot v+L_e^2 R \dot R-L_e^2 R \dot v+R^
3 \dot v) (2 L_e^2 m \dot v-L_e^2 R \dot R-L_e^2 R \dot v+R^3 \dot v)- \nb \\
& &2 L_e^4 R^2 \dot R^2] ((2 m-R) L_e^2+R^ 3)^{-1}
(L_e^2 m-R^3) L_e^{-4} R^{-3}+\dot R \ddot v- \ddot R \dot v
\lb{Ktautau+}
\eqn
\bq
K^{e}_{\theta\theta}= -\dot v((2 m-R) L_e^2+R^3) L_e^{-2}
\lb{Kthetatheta+}
\eq
\bq
K^{e}_{\phi\phi}=K^{e}_{\theta\theta}\sin^2(\theta).
\lb{Kphiphi+}
\eq
Since \cite{Lake}
\bq
[K_{\theta\theta}]= K^{e}_{\theta\theta}-K^{i}_{\theta\theta} = - M,
\lb{M}
\eq
where $M$ is the mass of the shell, we find that
\bq
M=\dot v R\left[ 1- \frac{2m}{R}-\left(\frac{R}{L_e}\right)^2 \right]+\dot t(1-2a R^2) R.
\lb{M1}
\eq
Then, substituting equations (\ref{dott2}) and (\ref{dotv2}) into (\ref{M1}) 
we get
\bq
M=-R\left[1-\frac{2m}{R} -\left(\frac{R}{L_e}\right)^2 + \dot R^2 \right]^{1/2} +
R\frac{ \left( 1 -2a R^2 + \dot R^2  \right)^{1/2}}
{(1-2a R^2)^{-(3\omega+2)/2}}.
\lb{M2}
\eq
In order to keep the ideas of MM as much as possible, we consider the thin 
shell as consisting
of a fluid with the equation of state, $p=(1-\gamma)\sigma$, where $\sigma$ and $p$ denote, 
respectively, the surface energy density and pressure of the shell and $\gamma$ is a constant. 
Then, the equation of motion of the shell is given by \cite{Lake}
\bq
\dot M + 8\pi R \dot R p = 4 \pi R^2 [T_{\alpha\beta}u^{\alpha}n^{\beta}]=
4\pi R^2 \left(T^e_{\alpha\beta}u_e^{\alpha}n_e^{\beta}-T^i_{\alpha\beta}u_i^{\alpha}n_i^{\beta} \right),
\lb{dotM}
\eq
where $u^{\alpha}$ is the four-velocity.
  
Let us first calculate $T^e_{\alpha\beta}u_e^{\alpha}n_e^{\beta}-T^i_{\alpha\beta}u_i^{\alpha}n_i^{\beta}$.  
We can get the energy-momentum tensor from reference \cite{Lobo} given by
\bq
T^i_{\alpha \beta}= (\rho+p_t) u^i_\alpha u^i_\beta + p_t g_{\alpha \beta} + (p_r -p_t)\chi_\alpha \chi_\beta,
\eq
where
$u^i_\alpha$ is the timelike four-velocity vector ($u^i_\alpha {u_i}^\alpha=-1$),
$\chi_\alpha$ is the unit spacelike vector in the radial direction ($\chi_\alpha \chi^\alpha =1$),
$n^i_\alpha$ is the spacelike normal vector $n^i_\alpha {n_i}^\alpha = {1 \over {f_1 f_2}}$ ($f_1>0$ and $f_2>0$),
in the region filled with the anisotropic fluid.
Thus, we have
\bq
u^i_\alpha \chi^\alpha=0,
\eq
and
\bq
n^i_\alpha {u_i}^\alpha=0.
\eq
Thus, since $T^e_{\alpha\beta}=0$, we can easily proof that 
$T^e_{\alpha\beta}u_e^{\alpha}n_e^{\beta}-T^i_{\alpha\beta}u_i^{\alpha}n_i^{\beta}=0$. 
Thus,
\bq
\dot M + 8\pi R \dot R (1-\gamma)\sigma = 0.
\lb{dotM1}
\eq
Recall that $\sigma = M/(4\pi R^2)$, we find that equation (\ref{dotM1}) has the solution
\bq
M=k R^{2(\gamma-1)},
\lb{Mk}
\eq
where $k$ is an integration constant. Substituting  equation (\ref{Mk}) into equation (\ref{M2}),
and rescaling $m, \; L_e \; a$ and $R$ as,
\bqn
m &\rightarrow& mk^{-\frac{1}{2\gamma-3}},\nb\\
a &\rightarrow& a k^{\frac{2}{2\gamma-3}},\nb\\
L_e &\rightarrow& L_e k^{\frac{2}{2\gamma-3}},\nb \\
R &\rightarrow& R k^{-\frac{1}{2\gamma-3}},
\eqn
we find that it can be written in the form of  equation (\ref{1.4}) with $a$ replaced by $R$, and
\bqn
& &V(R,m,L_e,\omega,a,\gamma)= \nb \\
& & -\frac{1}{2R^2 L_e^2 (b^2-1)} \left[2 R^{2 \gamma-2} b L_e \left(\frac{1}{b^2-1}\right) \left(R^{2 \gamma-2} b L_e- \right. \right. \nb \\
& &\left. \left. \sqrt{2 m L_e^2 R+R^4-2 R^4 L_e^2 a+L_e^2 R^{4 \gamma-4}-2 b^2 m L_e^2 R-b^2 R^4+2 R^4 L_e^2 a b^2} \right) \right.\nb \\
& & \left. +R^2 L_e^2-R^2 L_e^2 b^2+2 R^4 L_e^2 a b^2-2 m L_e^2 R-R^4-L_e^2 R^{4 \gamma-4} \right]
\lb{VR}
\eqn
where
\bq
\lb{b1}
b \equiv (1-2a R^2)^{-(1+3\omega/2)}.
\eq
Clearly, for any given constants $m$, $\omega$, $a$ and $\gamma$, equation (\ref{VR}) uniquely 
determines the collapse of the prototype  gravastar. Depending on the initial value $R_{0}$,  
the collapse can form either a black hole,  a gravastar,   a de Sitter, or a spacetime filled with
anisotropic dark energy fluid. In the last case, the thin shell
first collapses to a finite non-zero minimal radius and then expands to infinity.  

The exterior horizons are given by \cite{Shankaranarayanan}
\bq
r_{bh}= \frac{2m}{\sqrt{3 y}} \cos \left( \frac{\pi+\psi}{3} \right),
\label{rbh}
\eq
\bq
r_c= \frac{2m}{\sqrt{3 y}} \cos \left( \frac{\pi-\psi}{3} \right),
\eq
where $y=(m/L_e)^2$, $\psi= \arccos \left( 3\sqrt{3 y} \right)$, $r_{bh}$ denotes the black 
hole horizon and $r_c$ denotes the cosmological horizon. 

To  guarantee
that initially the spacetime does not have any kind of horizons,  cosmological or event,
we must restrict $R_{0}$ to the range,
\bq
\lb{2.2b}
r_{bh} < R_{0} < r_h \; or \; r_c,
\eq
where $R_0$ is the initial collapse radius. When $m = 0= a$, the thin shell disappears,
and the whole spacetime is Minkowski. So, in the following we shall not consider this case.

Since the potential  (\ref{VR}) is so complicated, it is too difficult to study it
analytically. Instead, in the following we shall study it numerically.

\section{Classifications of Matter, Dark Energy, and Phantom Energy  for Anisotropic Fluids}

Recently \cite{Chan08}, the classification of matter, dark and phantom energy 
for an anisotropic fluid in terms of the energy conditions was
studied since the pressure components 
may play very important roles and  can have quite different contributions.
In this paper, we will use this classification to study the collapse of the
dynamical prototype gravastars, constructed in the last section. 
The denomination used for each kind of fluid is given in Table 1.

\begin{table}
\caption{\label{tab:table1} This table summarizes the  classification of
the interior matter field, based on the energy conditions \cite{HE73}, where
we assume that $\rho \ge 0$.}
\begin{ruledtabular}
\begin{tabular}{cccc}
Matter & Condition 1 & Condition 2  & Condition 3 \\
\hline
Standard Matter           & $\rho+p_r+2p_t\ge 0$ & $\rho+p_r\ge 0$ & $\rho+p_t\ge 0$ \\
\hline
Dark Energy               & $\rho+p_r+2p_t <  0$ & $\rho+p_r\ge 0$ & $\rho+p_t\ge 0$ \\
\hline
                          &                      & $\rho+p_r <  0$ & $\rho+p_t\ge 0$ \\
Repulsive Phantom Energy  & $\rho+p_r+2p_t <  0$ & $\rho+p_r\ge 0$ & $\rho+p_t <  0$ \\
                          &                      & $\rho+p_r <  0$ & $\rho+p_t <  0$ \\
\hline
                          &                      & $\rho+p_r <  0$ & $\rho+p_t\ge 0$ \\
Attractive Phantom Energy & $\rho+p_r+2p_t\ge 0$ & $\rho+p_r\ge 0$ & $\rho+p_t <  0$ \\
                          &                      & $\rho+p_r <  0$ & $\rho+p_t <  0$ \\
\end{tabular}
\end{ruledtabular}
\end{table}

\begin{table}
\caption{\label{tab:table2} This table summarizes the  classification of matter
on the thin shell, based on the energy conditions \cite{HE73}. The last column indicates
the particular values of the parameter $\gamma$, where we assume that $\rho \ge 0$.}
\begin{ruledtabular}
\begin{tabular}{cccc}
Matter & Condition 1 & Condition 2  & $\gamma$ \\
\hline
Standard Matter           & $\sigma+2p\ge 0$ & $\sigma+p\ge 0$ & -1 or 0  \\
Dark Energy               & $\sigma+2p <  0$ & $\sigma+p\ge 0$ &  7/4 \\
Repulsive Phantom Energy  & $\sigma+2p <  0$ & $\sigma+p <  0$ &   3  \\
\end{tabular}
\end{ruledtabular}
\end{table}

In order to consider the equations (\ref{ds2-}) and (\ref{prpt}) for describing dark energy
stars we must analyze carefully the ranges of the parameter $\omega$ that in
fact furnish the expected fluids.  It can be shown that the
condition $\rho+p_r>0$ is violated for $\omega<-1$ and fulfilled for $\omega>-1$, 
for any values of $R$ and $b$.
The conditions $\rho+p_t>0$ and $\rho+p_r+2p_t>0$ are satisfied for $\omega<-1$
and $-1/3<\omega<0$, for any values of $R$ and $b$.  
For the other intervals of
$\omega$  the
 energy conditions depend
on very complicated relations of $R$ and $b$ \cite{Chan08}.
This provides an explicit example, in which the definition of dark energy must 
be dealed with great care.  Another case was provided in a previous work 
\cite{Chan08}.  

In order to fulfill the energy condition $\sigma+2p\ge0$ of the shell
and assuming that
$p=(1-\gamma)\sigma$ we must have $\gamma \le 3/2$. On the other hand, in order
to satisfy the condition $\sigma+p\ge 0$, we obtain $\gamma \le 2$.
Hereinafter, we will use only some particular values of the parameter
$\gamma$ which are analyzed in this work. See Table II.

In the next sections we will discuss the different possibilities for the
type of system that can be formed from the study of the potential
$V(R,m,L_e,\omega,a,\gamma)$: 
black hole of standard matter or dark energy, stable or "bounded excursion" 
gravastar and even naked singularity.

\section{Structures Formed}

Here we can find many types of systems, depending on the combination of the 
constitution matter of the shell and core.  Among them, there are formation of
black holes, stable and "bounded excursion" gravastars, as it has already 
shown in our previous works \cite{JCAP}-\cite{JCAP4}, 
and even a naked singularity constituted exclusively of dark energy. 
All of them are  listed in the table III. 

As can be seen in the figure \ref{fig11}, depending on the value of the cosmological constant, we can see that $V(R) = 0$ now can have one, two  or three real roots. Then, we 
have, say, $R_{i}$, where $R_{i+1} > R_{i}$. For $L_e=L_1$ (corresponding to $\Lambda=0$)  If we choose $R_{0} > R_{3}$ none structure is allowed in this region because the potential is greater than the zero.  However, if we choose $R_{2} < R_{0} < R_{3}$, the collapse will bounce back and forth between $R = R_{1}$ and $R = R_{2}$. Such a  possibility is better shown in the figure \ref{fig13}. This is exactly the so-called "bounded excursion" model mentioned in \cite{VW04}, and studied in some details in \cite{JCAP}-\cite{JCAP4}.  Of 
course, in a realistic situation, the star will emit both gravitational waves 
and particles, and the potential will be self-adjusted to produce a minimum at 
$R = R_{static}$ where $V\left(R=R_{static}\right) = 0 = V'\left(R=R_{static}\right)$ 
whereby a gravastar is finally formed \cite{VW04,JCAP,JCAP1,JCAP2}. For $R_{0} < R_{1}$ a black hole is formed in the end of the collapse of the shell. 

The scenario above can significantly be changed if we consider $\Lambda>0$. 
In this case for $L_e>L_c$, we also have bounded excursion gravastars if 
$R_{2} < R_{0} < R_{3}$. However, for $R_{0} < R_{1}$ the final structure 
can be now a black hole or a naked singularity since the presence of the 
cosmological constant above a certain limit ($L_e^*$) eliminates the 
event horizon (its radius becomes negative), as can be seen in the table IV. 
See figures \ref{fig11a}, \ref{fig11b} and \ref{fig11c}. This is the first 
evidence of a naked singularity formation from a gravastar model. Moreover 
for $L_e=L_c$, then $R_{2} = R_{3}$, a stable gravastar is formed if 
$R_{0} = R_{2}$, while for $L_e<L_c$ there is only one real root. Note that 
for any value of $L_e>L_e^*$, a naked singularity is formed for small 
initial radius of the shell.

Another two very interesting structures that arise from this model are a 
black hole, represented  by figures \ref{fig11d}, \ref{fig14}, \ref{fig20} 
and \ref{fig21}, 
and a stable gravastar, represented by figure \ref{fig18}, all of them made 
of dark energy. It is remarkable that the stable gravastar is constituted 
only by phantom energy. It means that a system 
constituted only by dark energy is able to achieve an equilibrium state or 
even to collapse. 
Thus, solving equation (\ref{M2}) for $\dot R(\tau)$ we can integrate 
$\dot R(\tau)$ and obtain $R(\tau)$, which are
shown in the figures \ref{fig11a2}, \ref{fig12}, \ref{fig15} and \ref{fig16} for the case F.
 
\begin{table}
\caption{\label{tab:table3}This table summarizes all possible kind of energy
of the interior fluid and of the shell and compares the formed structures
in the two gravastar models ($\Lambda=0$, $\Lambda>0$). The letters SG, UG, BEG, BH, NS and N
denote stable gravastar, unstable gravastar, bounded excursion gravastar,
black hole, naked singularity and none, respectively.}
\begin{ruledtabular}
\begin{tabular}{cccccc}
Case & Interior Energy & Shell Energy & Figures & Structures ($\Lambda=0$) & Structures ($\Lambda>0$)\\
\hline
A & Standard           & Standard           &  & BH & BH \\
B & Standard           & Dark               &  & BH & BH \\
C & Standard           & Repulsive Phantom  &  & UG/BH & BH \\
D & Dark               & Standard           &  & SG & N\\
E & Dark               & Dark               & \ref{fig20} & UG/BH & BH\\
F & Dark               & Repulsive Phantom  & \ref{fig11},\ref{fig11a1},\ref{fig11a},\ref{fig11b},\ref{fig11c},\ref{fig11d},\ref{fig13} & BH/BEG &  BEG/SG/NS\\
  &                    &                    & \ref{fig14},\ref{fig17} & BH/UG &  BH\\
G & Repulsive Phantom  & Standard           &  & BH & BH\\
H & Repulsive Phantom  & Dark               & \ref{fig21} & BH & BH\\
I & Repulsive Phantom  & Repulsive Phantom  & \ref{fig18} & SG & N\\
J & Attractive Phantom & Standard           &  & N & N \\
K & Attractive Phantom & Dark               &  & N & N \\
L & Attractive Phantom & Repulsive Phantom  &  & N & N \\
\end{tabular}
\end{ruledtabular}
\end{table}

\begin{table}
\caption{\label{tab:table4} Radii $r_{bh}$ and $r_{c}$ calculated using
equation (\ref{rbh}) as a function of $L_e$ 
for $L_e < L_{c}$ and for $L_e > L_{c}$ where $m=10^{-9}$ and $L_{c}=61.49177124$.}
\begin{ruledtabular}
\begin{tabular}{c|cc} 
\hline
$r_{bh}$ & $r_{c}$ & $L_e$ \\ 
\hline
$ 0.2072568092 \times 10^{-8}$ & $1.0$  & $1.0 $ \\
$ 0.4589337770 \times 10^{-8}$ & $5.0$  & $5.0 $ \\
$ 0.9178675538 \times 10^{-8}$ & $10.0$ & $10.0$ \\
$			     $ & $    $ & $L_e^*$ \\
$-0.4736659684 \times 10^{-8}$ & $20.0$ & $20.0$ \\
$-0.7104989528 \times 10^{-8}$ & $30.0$ & $30.0$ \\
$-0.9473319370 \times 10^{-7}$ & $40.0$ & $40.0$ \\
$-0.1184164921 \times 10^{-7}$ & $50.0$ & $50.0$ \\
$-0.1456327969 \times 10^{-7}$ & $61.49177126  $ & $L_{c}$ \\
$-0.1657830889 \times 10^{-7}$ & $70.0$ & $70.0$ \\
$-0.1894663874 \times 10^{-7}$ & $80.0$ & $80.0$ \\
$-0.2131496858 \times 10^{-7}$ & $90.0$ & $90.0$ \\
$-0.2368329842 \times 10^{-7}$ & $1.0 \times 10^{2} $ & $1.0 \times 10^{2}$ \\
$-0.2368329842 \times 10^{-6}$ & $1.0 \times 10^{3} $ & $1.0 \times 10^{3}$ \\
$-0.2368329842 \times 10^{-5}$ & $1.0 \times 10^{4} $ & $1.0 \times 10^{4}$ \\
$-0.2368329842 \times 10^{-4}$ & $1.0 \times 10^{5} $ & $1.0 \times 10^{5}$ \\
$-0.2368329842 \times 10^{-3}$ & $1.0 \times 10^{6} $ & $1.0 \times 10^{6}$ \\
$-0.2368329842 \times 10^{-2}$ & $1.0 \times 10^{7} $ & $1.0 \times 10^{7}$ \\
$-0.2368329842 \times 10^{-1}$ & $1.0 \times 10^{8} $ & $1.0 \times 10^{8}$ \\
$-2.368329842                $ & $1.0 \times 10^{10}$ & $1.0 \times 10^{10}$ \\
$-2.368329842 \times 10^{1}  $ & $1.0 \times 10^{11}$ & $1.0 \times 10^{11}$ \\
$-2.368329842 \times 10^{2}  $ & $1.0 \times 10^{12}$ & $1.0 \times 10^{12}$ \\
$-0.2368329842 \times 10^{13}$ & $1.0 \times 10^{22}$ & $1.0 \times 10^{22}$ \\
$-0.2368329842 \times 10^{81}$ & $1.0 \times 10^{90}$ & $1.0 \times 10^{90}$ \\
$-\infty$ & $\infty$ & $\infty$ \\
\hline
\end{tabular}
\end{ruledtabular}
\end{table}
                                             
\section{Conclusions}

In this paper, we have studied the problem of the stability of gravastars by
constructing dynamical three-layer models  of VW \cite{VW04},
which consists of an internal anisotropic dark energy fluid, a dynamical infinitely thin  shell of
perfect fluid with the equation of state $p = (1-\gamma)\sigma$, and an external 
de Sitter-Schwarzschild spacetime.

We have shown explicitly that the final output can be a black
hole, a "bounded excursion" stable gravastar depending on the total mass 
$m$ of the system, the cosmological constant $L_e$, the parameter $\omega$, 
the constant $a$, the parameter $\gamma$ and
the initial position $R_{0}$ of the dynamical shell. All these possibilities
have non-zero measurements in the phase space of $m$, $L_e$, $a$, $\omega$, $\gamma$ 
and $R_{0}$.  All the results can be summarized in Table III. 

An interesting result that we can deduce from Table III is that we can have
black hole and stable gravastar formation even with an interior and a shell constituted of dark
and repulsive dark energy (cases E, F, H and I).

\begin{figure}
\vspace{.2in}
\centerline{\psfig{figure=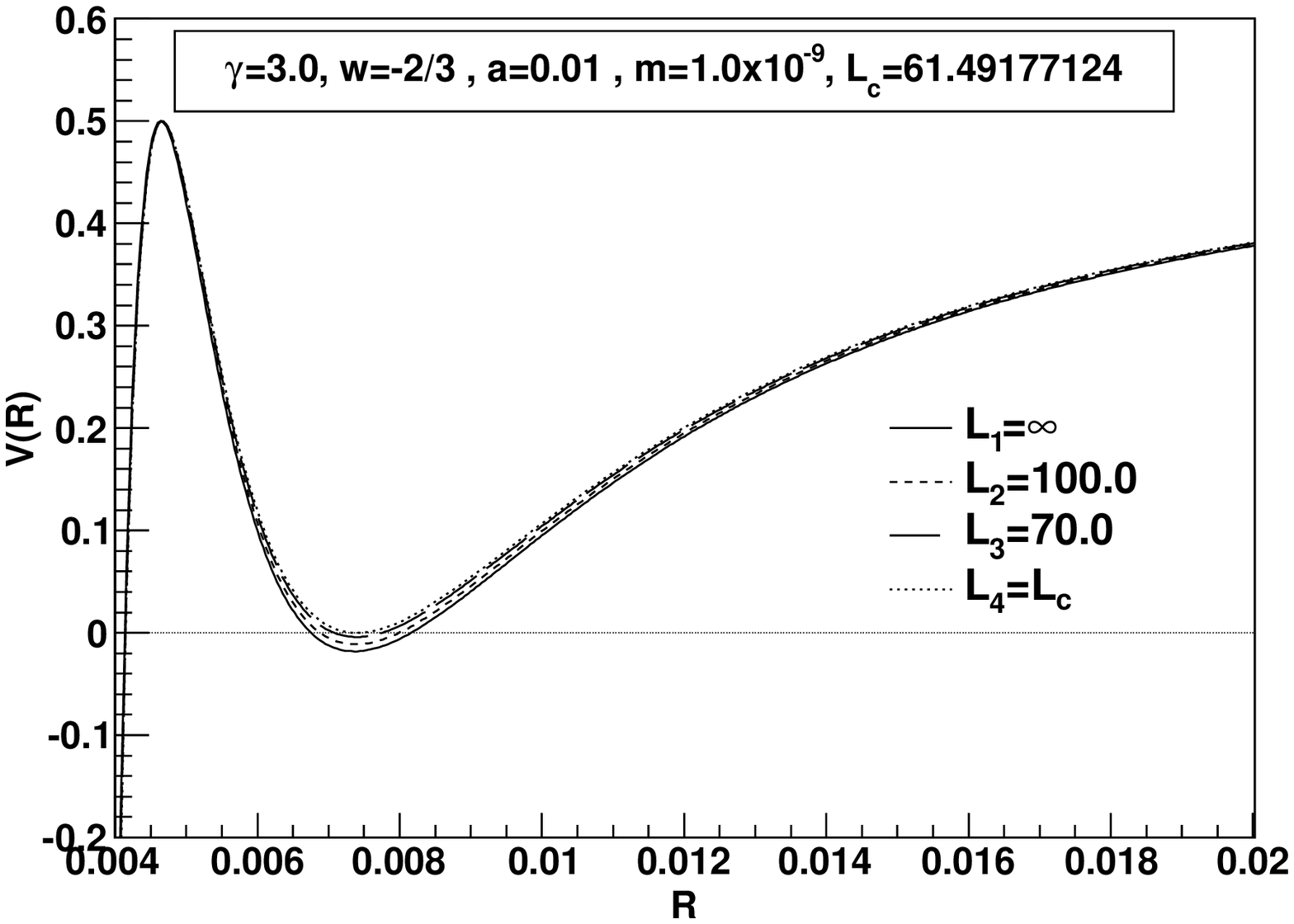,width=3.3truein,height=3.0truein}\hskip
.25in \psfig{figure=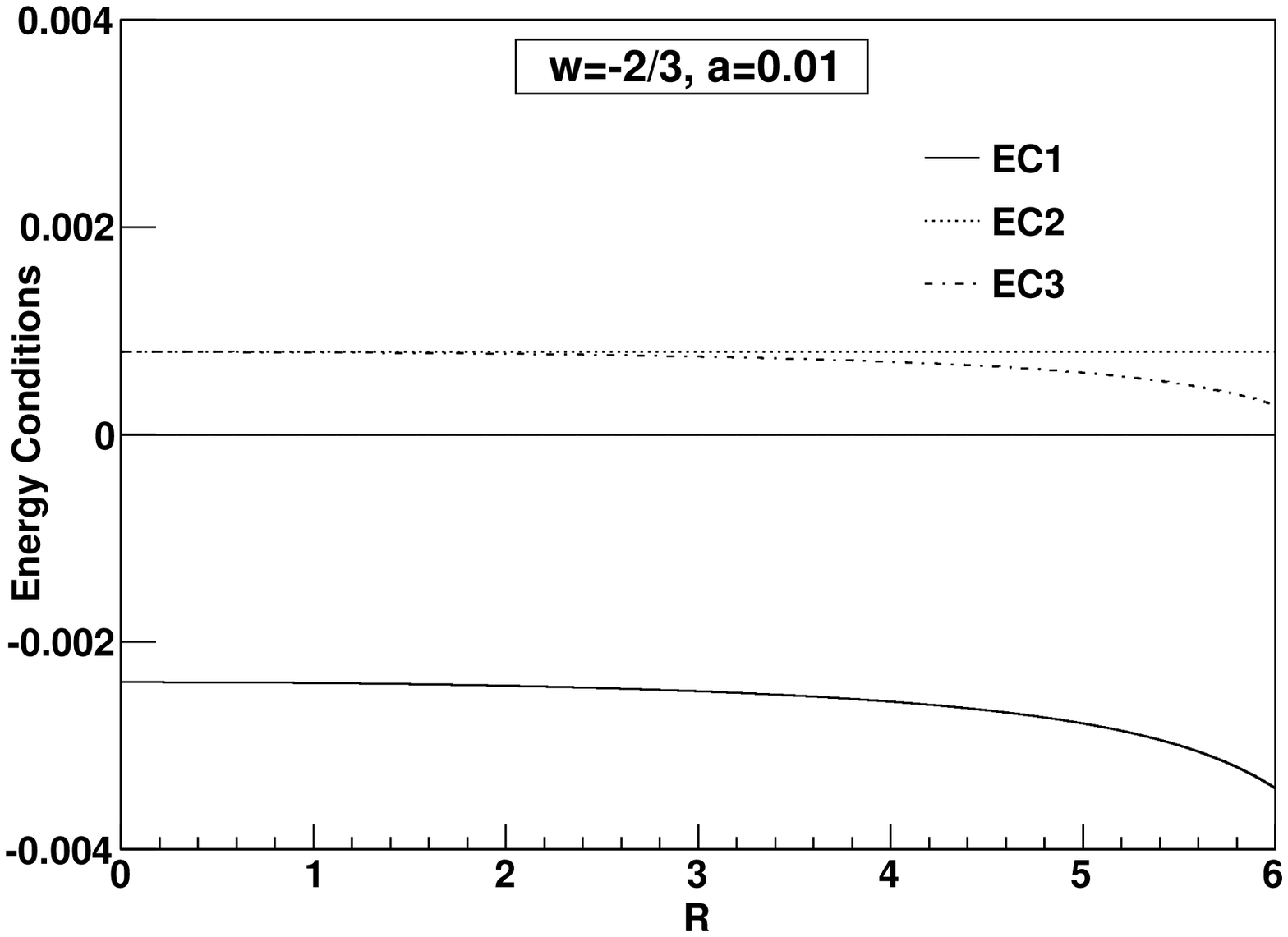,width=3.3truein,height=3.0truein}
\hskip .5in} \caption{The potential $V(R)$ and the energy conditions EC1$\equiv \rho+p_r+2p_t$, 
EC2$\equiv \rho+p_r$ and EC3$\equiv \rho+p_t$, for $\gamma=3$,
$\omega=-2/3$, $a=0.01$ and $m_c=0.1054688609\times 10^{-8}$. The horizons are: $r_h=7.071067814$; $r_{bh}=-0.2368329842\times 10^{-7}$, $r_c=99.99999998$ ($L_e=100$); $r_{bh}=-0.1657830889\times 10^{-7}$, $r_c=70$ ($L_e=70$); $r_{bh}=-0.1456327969\times 10^{-7}$, $r_c=61.49177126$ ($L_e=61.49177124$). Note that we have a gravastar enclosing a naked singularity.  Then, for small $R$, the shell can collapse to form  a naked singularity. {\bf Case F}}
\label{fig11}
\end{figure}

We also would like to point out the significant influence of the presence of the
exterior cosmological constant to formation of this kind of structure. Note
that in the case I, represented by figure \ref{fig18}, we have a stable gravastar
(for $\Lambda=0$) or no formed structure (for $\Lambda > 0$), as well as in the case D, 
where we have a traditional stable gravastar only for $\Lambda = 0$. 
Still more interesting
is the case F, represented by figure \ref{fig11}, where for small radius of the shell we
have formation of a black hole (for $\Lambda=0$) or a naked singularity 
(for  $\Lambda > 0$). This is the first time in the literature that a naked 
singularity emerges from a gravastar model.  Besides, the figures \ref{fig11a2}, \ref{fig12},
\ref{fig15} and \ref{fig16} give us examples of the dynamical evolution of a gravastar to a naked singularity, 
a "bounded excursion" gravastar, a black hole and an unstable gravastar, respectively.

\begin{figure}
\vspace{.2in}
\centerline{\psfig{figure=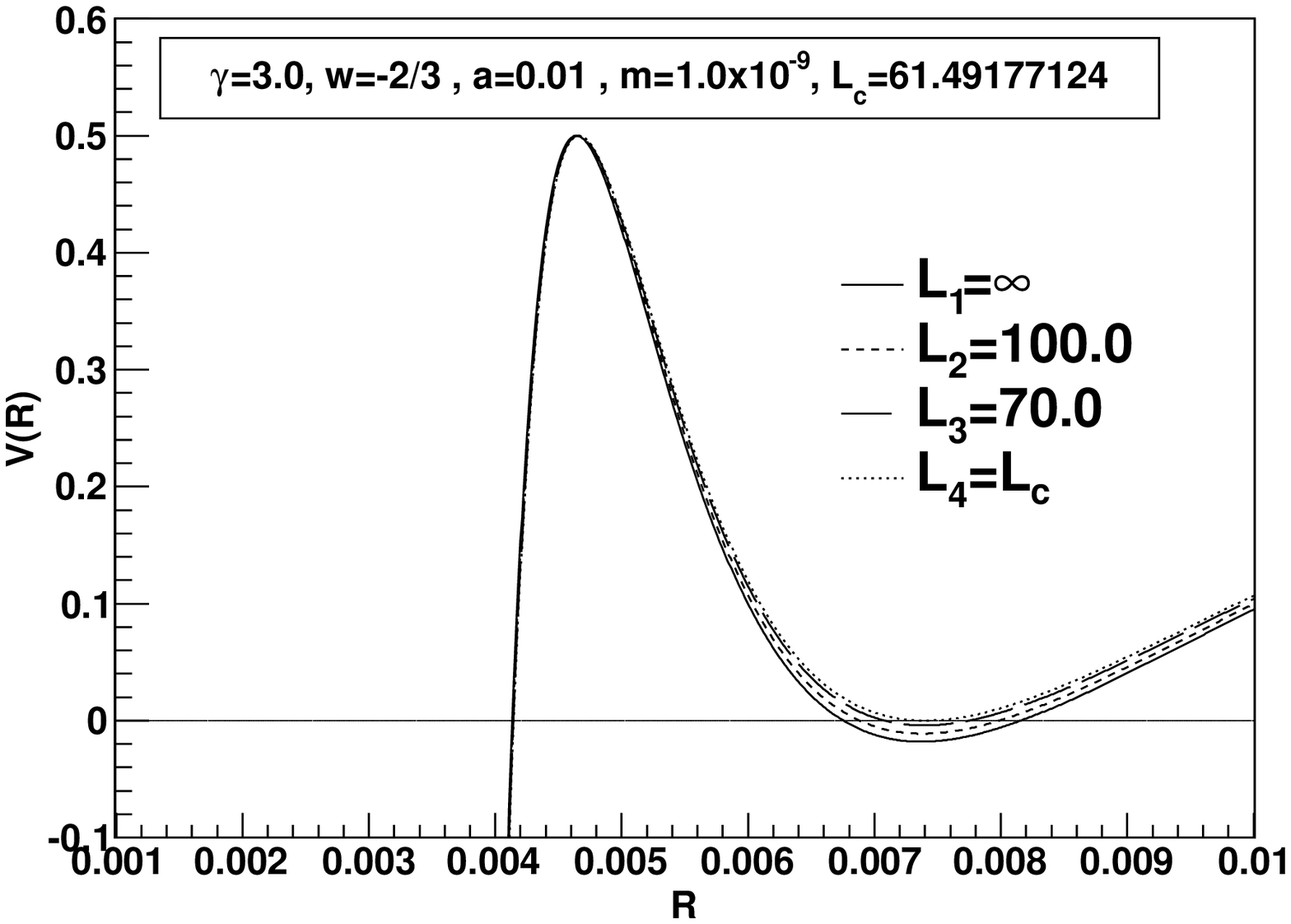,width=3.3truein,height=3.0truein}\hskip
.25in \psfig{figure=CondEnwm2div3a0v01.eps,width=3.3truein,height=3.0truein}
\hskip .5in} \caption{The zoom of the potential $V(R)$ (figure \ref{fig11}) and the energy conditions EC1$\equiv \rho+p_r+2p_t$, 
EC2$\equiv \rho+p_r$ and EC3$\equiv \rho+p_t$, for $\gamma=3$,
$\omega=-2/3$, $a=0.01$ and $m_c=0.1054688609\times 10^{-8}$. The horizons are: $r_h=7.071067814$; $r_{bh}=-0.2368329842\times 10^{-7}$, $r_c=99.99999998$ ($L_e=100$); $r_{bh}=-0.1657830889\times 10^{-7}$, $r_c=70$ ($L_e=70$); $r_{bh}=-0.1456327969\times 10^{-7}$, $r_c=61.49177126$ ($L_e=61.49177124$). Note that we have a gravastar enclosing a naked singularity.  Then, for small $R$, the shell can collapse to form  a naked singularity. {\bf Case F}}
\label{fig11a1}
\end{figure}

\begin{figure}
\vspace{.2in}
\centerline{\psfig{figure=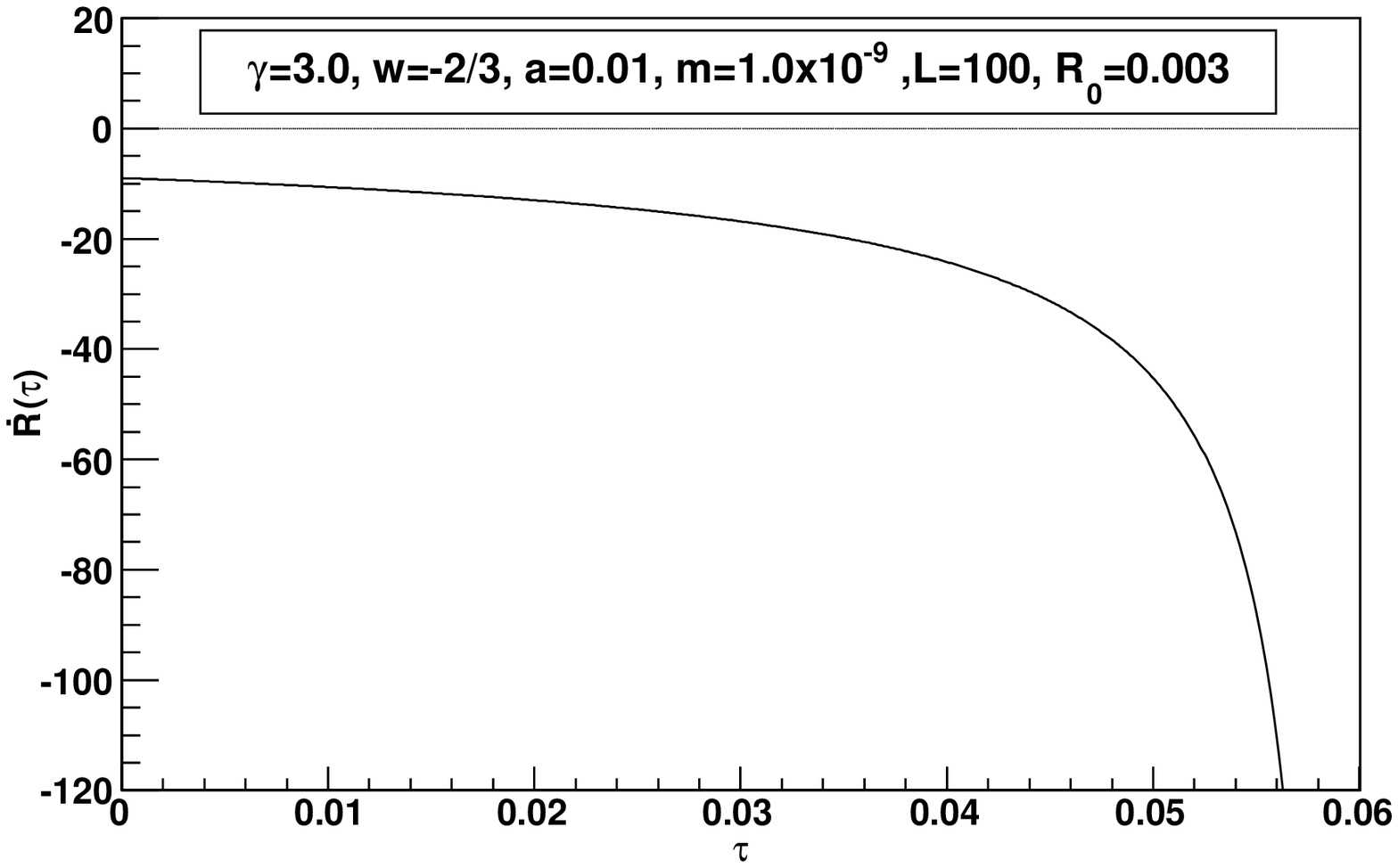,width=3.3truein,height=3.0truein}\hskip
.25in \psfig{figure=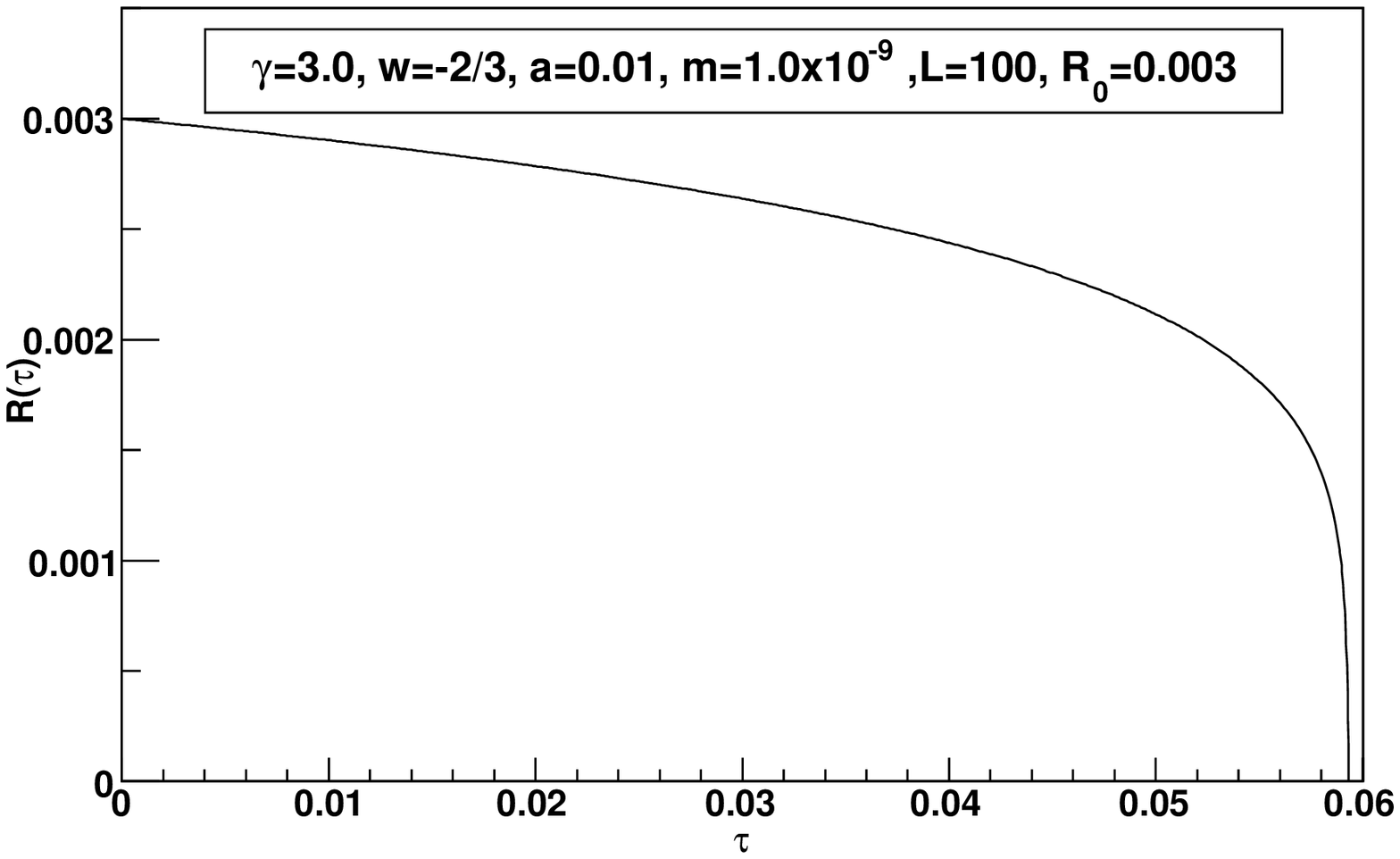,width=3.3truein,height=3.0truein}
\hskip .5in} \caption{These figures represent the dynamical evolution of a gravastar to a naked singularity for the potential given by the figure \ref{fig11a1},
assuming $R_0=R(0)=0.003$.}
\label{fig11a2}
\end{figure}

\begin{figure}
\vspace{.2in}
\centerline{\psfig{figure=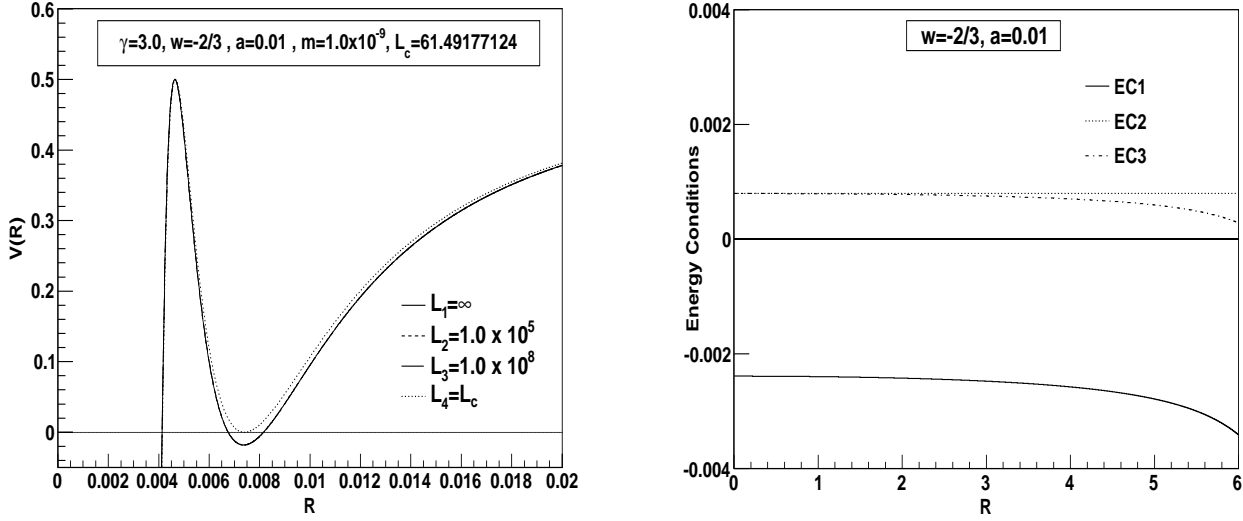,width=3.3truein,height=3.0truein}\hskip
.25in \psfig{figure=CondEnwm2div3a0v01.eps,width=3.3truein,height=3.0truein}
\hskip .5in} \caption{The zoom of the potential $V(R)$ (figure \ref{fig11}) and the energy conditions EC1$\equiv \rho+p_r+2p_t$, 
EC2$\equiv \rho+p_r$ and EC3$\equiv \rho+p_t$, for $\gamma=3$,
$\omega=-2/3$, $a=0.01$ and $m_c=0.1054688609\times 10^{-8}$. The horizons are: $r_h=7.071067814$; $r_{bh}=-0.2368329842 \times 10^{-4}$, $r_c=1.0 \times 10^{5}$ ($L_e=1.0 \times 10^{5}$); $r_{bh}=-0.2368329842 \times 10^{-1}$, $r_c=1.0 \times 10^{8}$ ($L_e=1.0 \times 10^{8}$); $r_{bh}=-0.1456327969\times 10^{-7}$, $r_c=61.49177126$ ($L_e=61.49177124$). Note that we have a gravastar enclosing a naked singularity.  Then, for small $R$, the shell can collapse to form  a naked singularity. {\bf Case F}}
\label{fig11a}
\end{figure}

\begin{figure}
\vspace{.2in}
\centerline{\psfig{figure=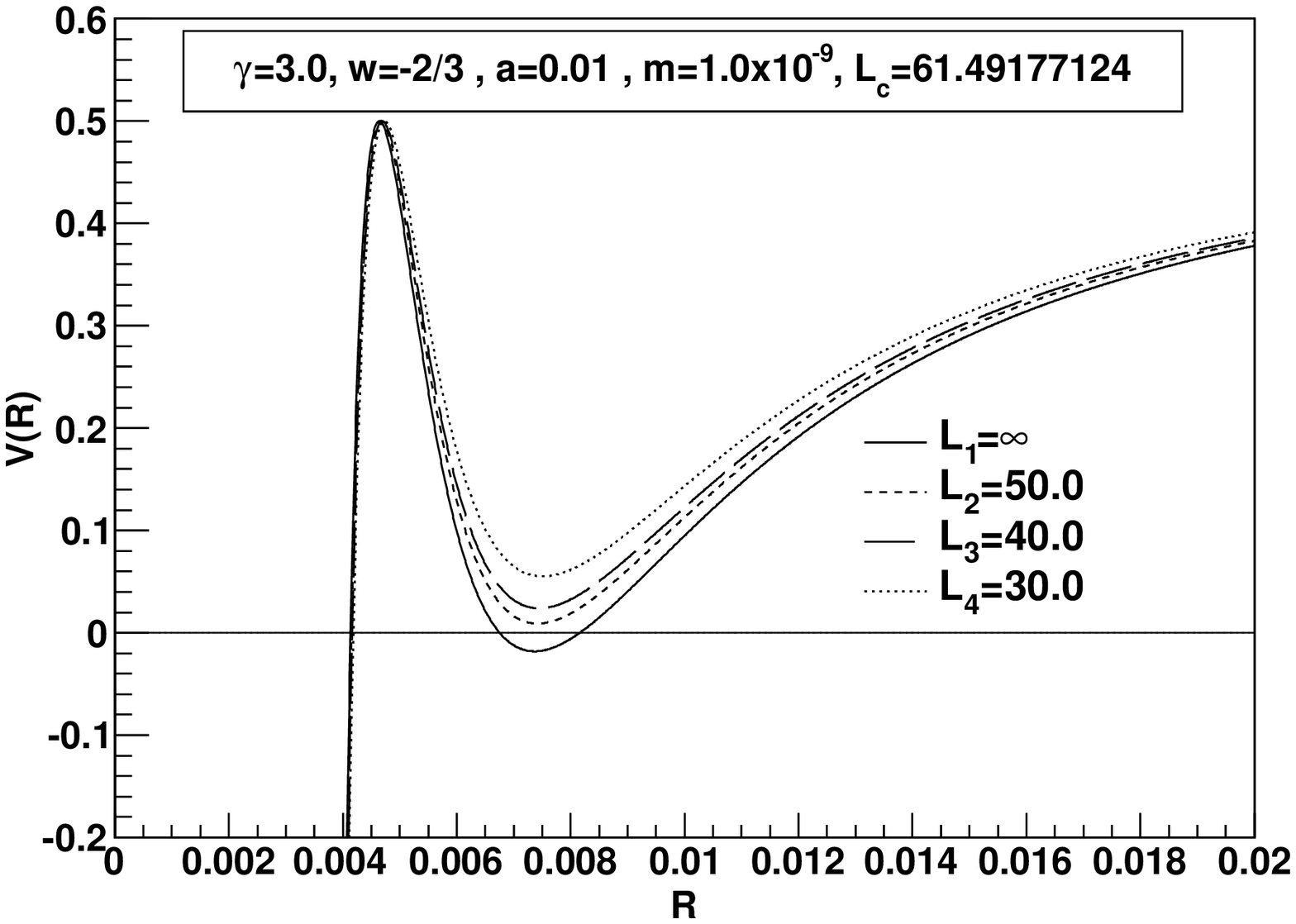,width=3.3truein,height=3.0truein}\hskip
.25in \psfig{figure=CondEnwm2div3a0v01.eps,width=3.3truein,height=3.0truein}
\hskip .5in} \caption{The zoom of the potential $V(R)$ (figure \ref{fig11}) and the energy conditions EC1$\equiv \rho+p_r+2p_t$, 
EC2$\equiv \rho+p_r$ and EC3$\equiv \rho+p_t$, for $\gamma=3$,
$\omega=-2/3$, $a=0.01$ and $m_c=0.1054688609\times 10^{-8}$. The horizons are: $r_h=7.071067814$; $r_{bh}=-0.1184164921 \times 10^{-7}$, $r_c=50.0$ ($L_e=50$); $r_{bh}=-0.9473319370 \times 10^{-7}$, $r_c=40$ ($L_e=40$); $r_{bh}=-0.7104989528 \times 10^{-8}$, $r_c=30.0$ ($L_e=30$). Note that we have a gravastar enclosing a naked singularity.  Then, for small $R$, the shell can collapse to form  a naked singularity. {\bf Case F}}
\label{fig11b}
\end{figure}

\begin{figure}
\vspace{.2in}
\centerline{\psfig{figure=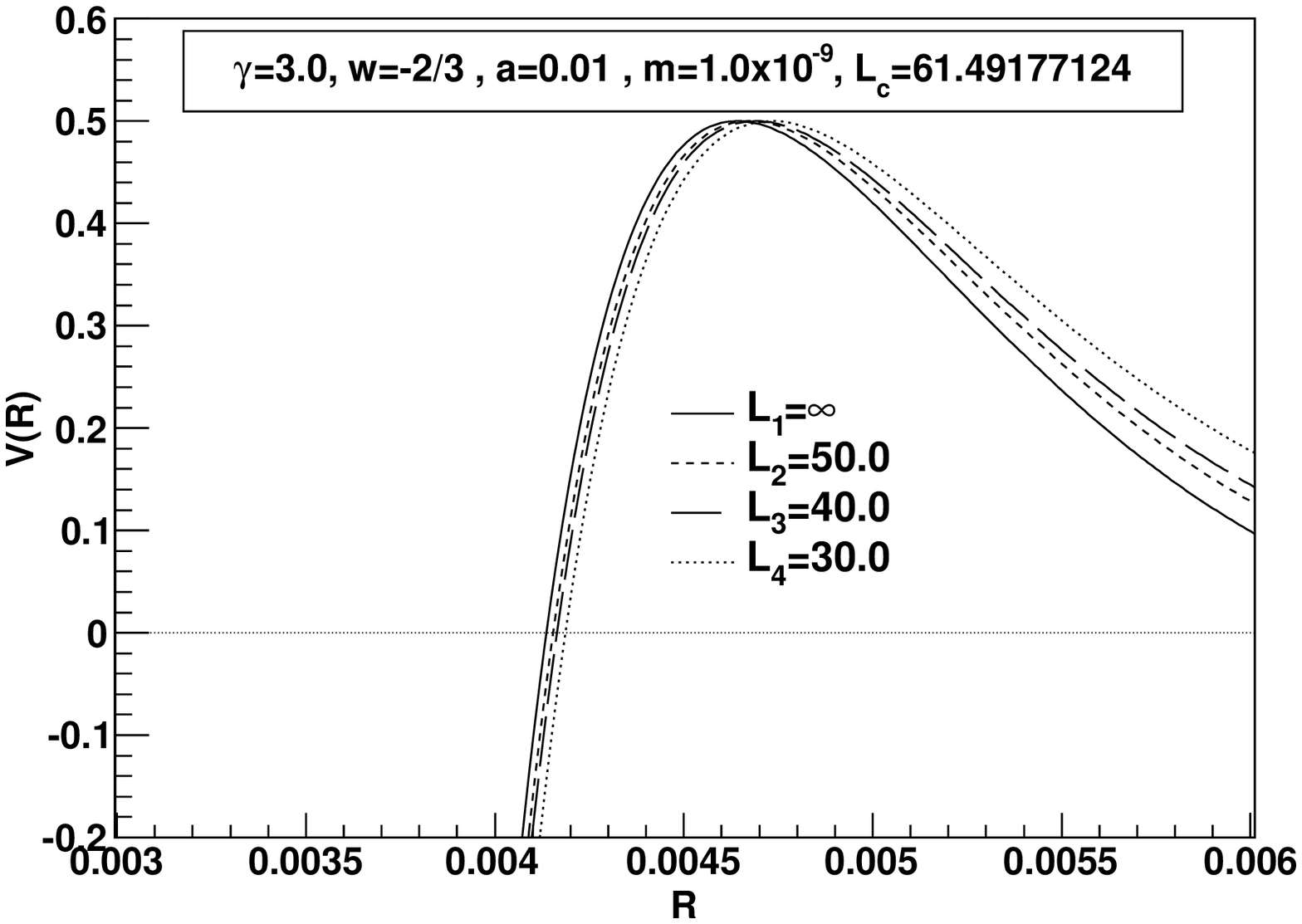,width=3.3truein,height=3.0truein}\hskip
.25in \psfig{figure=CondEnwm2div3a0v01.eps,width=3.3truein,height=3.0truein}
\hskip .5in} \caption{The zoom of the potential $V(R)$ (figure \ref{fig11}) and the energy conditions EC1$\equiv \rho+p_r+2p_t$, 
EC2$\equiv \rho+p_r$ and EC3$\equiv \rho+p_t$, for $\gamma=3$,
$\omega=-2/3$, $a=0.01$ and $m_c=0.1054688609\times 10^{-8}$. The horizons are: $r_h=7.071067814$; $r_{bh}=-0.1184164921 \times 10^{-7}$, $r_c=50.0$ ($L_e=50$); $r_{bh}=-0.9473319370 \times 10^{-7}$, $r_c=40$ ($L_e=40$); $r_{bh}=-0.7104989528 \times 10^{-8}$, $r_c=30.0$ ($L_e=30$). Note that we have a gravastar enclosing a naked singularity.  Then, for small $R$, the shell can collapse to form  a naked singularity. {\bf Case F}}
\label{fig11c}
\end{figure}

\begin{figure}
\vspace{.2in}
\centerline{\psfig{figure=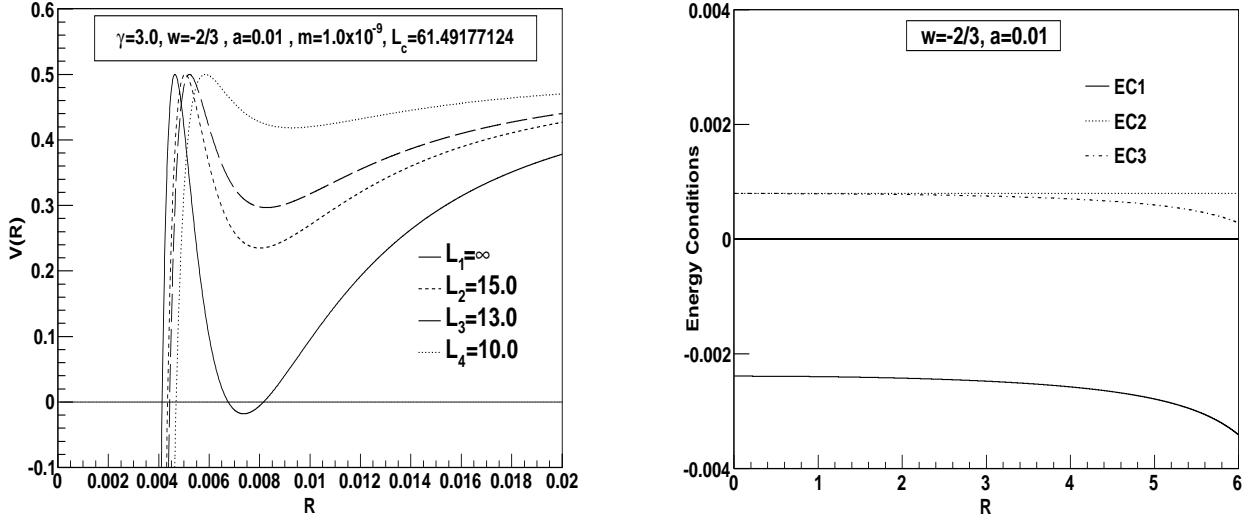,width=3.3truein,height=3.0truein}\hskip
.25in \psfig{figure=CondEnwm2div3a0v01.eps,width=3.3truein,height=3.0truein}
\hskip .5in} \caption{The zoom of the potential $V(R)$ (figure \ref{fig11}) and the energy conditions EC1$\equiv \rho+p_r+2p_t$, 
EC2$\equiv \rho+p_r$ and EC3$\equiv \rho+p_t$, for $\gamma=3$,
$\omega=-2/3$, $a=0.01$ and $m_c=0.1054688609\times 10^{-8}$. The horizons are: $r_h=7.071067814$; $r_{bh}=0.9178675538 \times 10^{-8}$, $r_c=10.0$ ($L_e=10$). Note that we have a gravastar enclosing a black hole.  Then, for small $R$, the shell can collapse to form  a black hole. {\bf Case F}}
\label{fig11d}
\end{figure}

\begin{figure}
\vspace{.2in}
\centerline{\psfig{figure=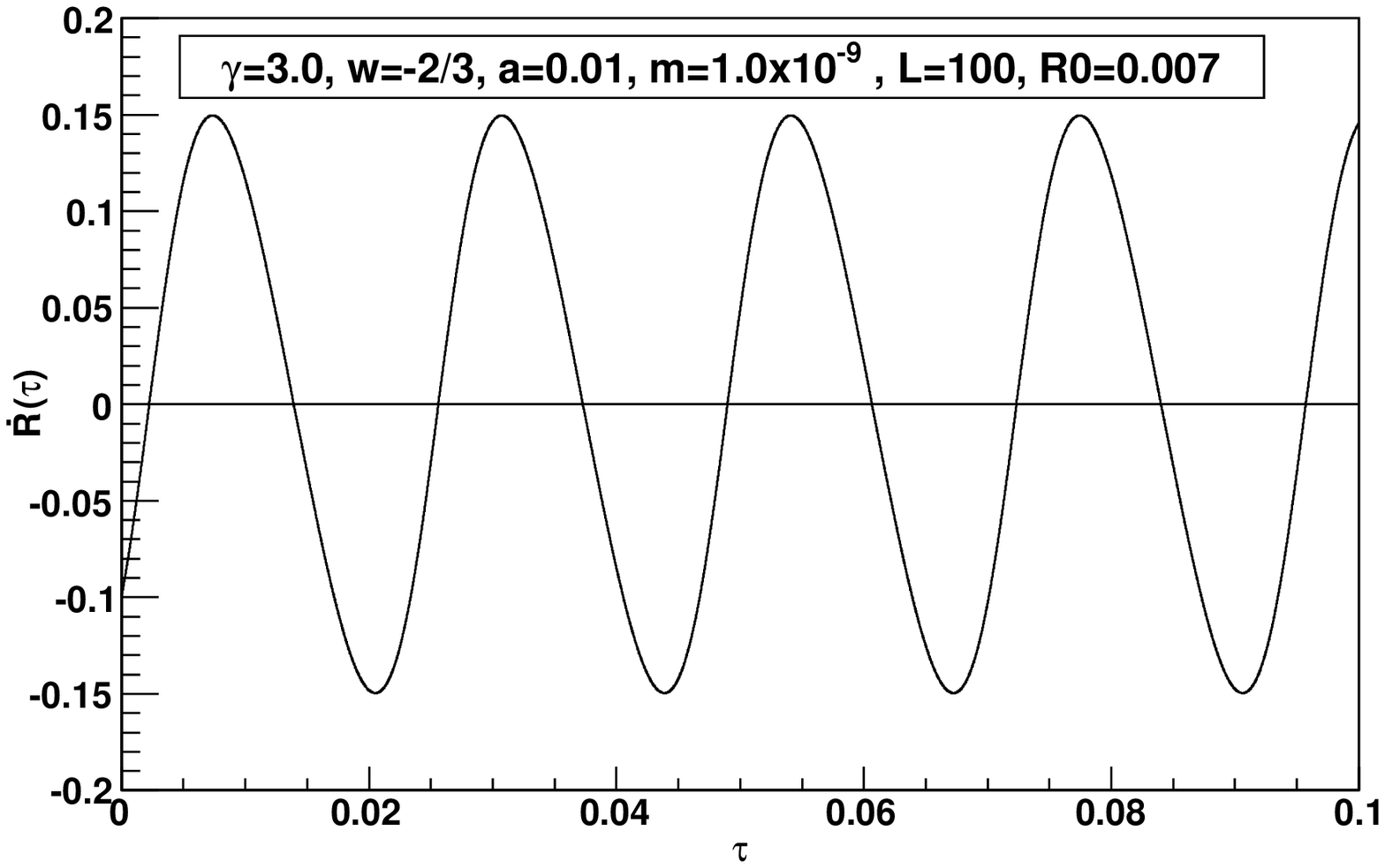,width=3.3truein,height=3.0truein}\hskip
.25in \psfig{figure=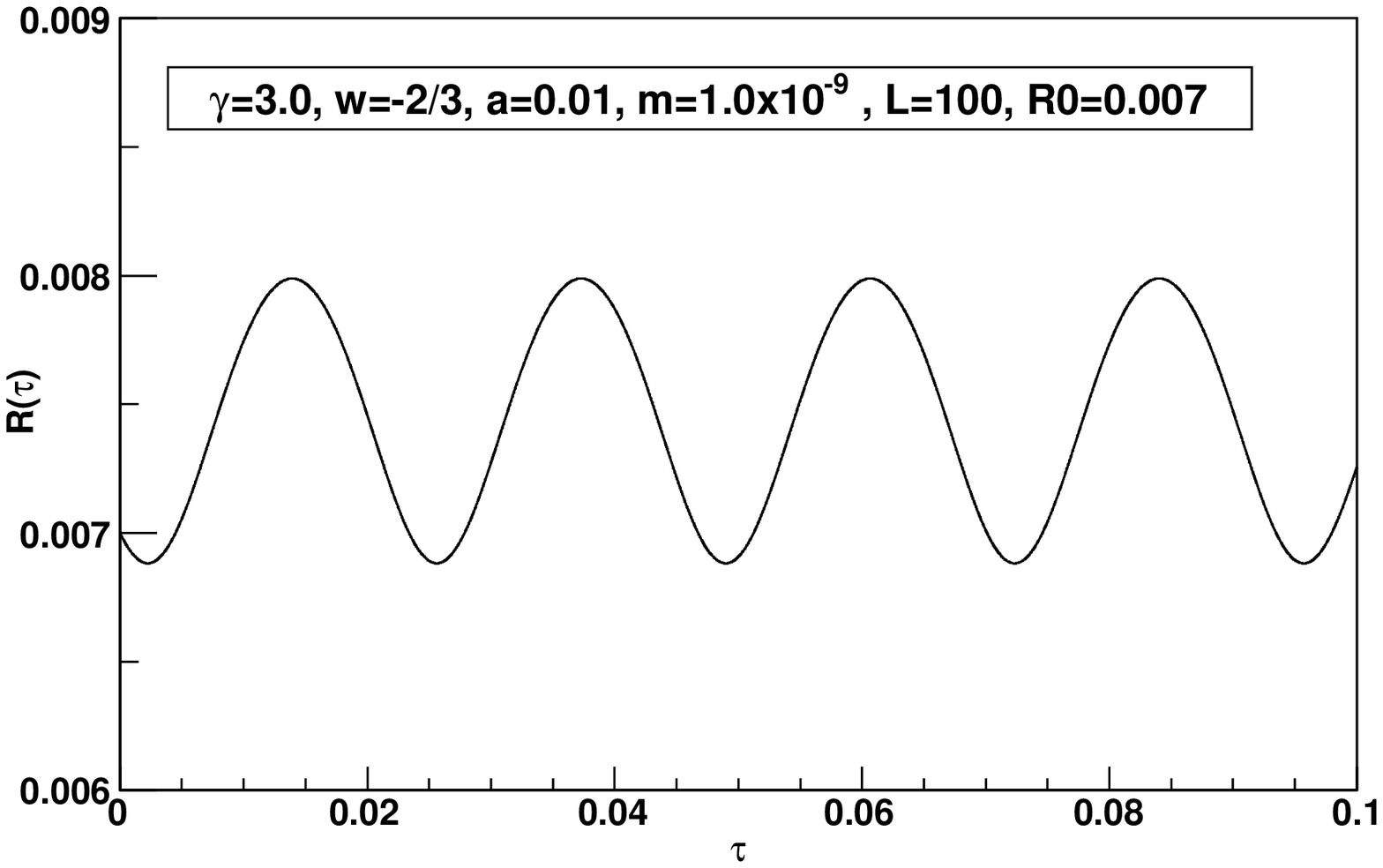,width=3.3truein,height=3.0truein}
\hskip .5in} \caption{These figures represent the dynamical evolution of a "bounded excursion" gravastar with the potential given by the figure \ref{fig11},
assuming $R_0=R(0)=0.007$.}
\label{fig12}
\end{figure}

\begin{figure}
\vspace{.2in}
\centerline{\psfig{figure=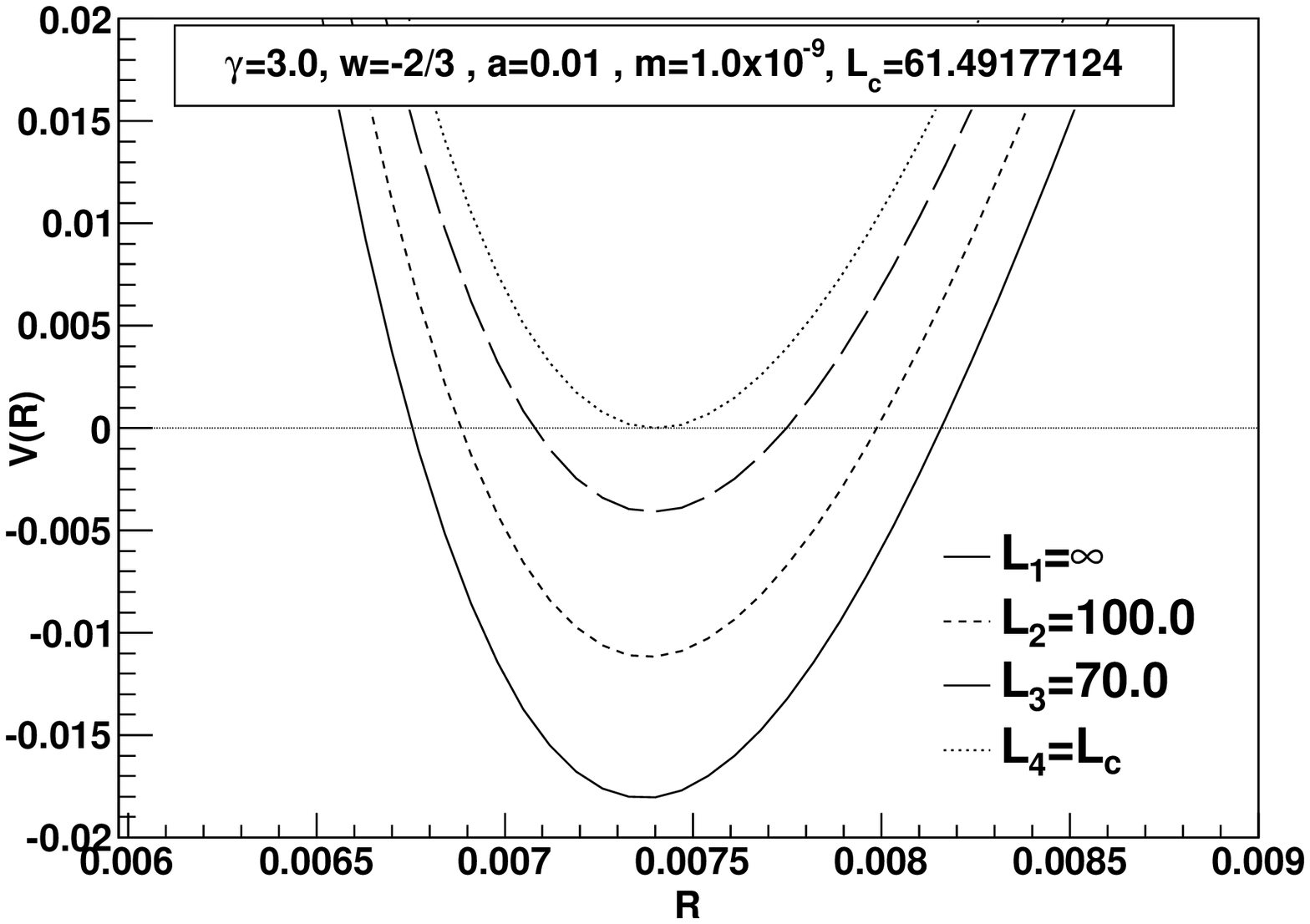,width=3.3truein,height=3.0truein}\hskip
.25in \psfig{figure=CondEnwm2div3a0v01.eps,width=3.3truein,height=3.0truein}
\hskip .5in} \caption{Zoom of the potential $V(R)$ of figure \ref{fig11} and the energy conditions EC1$\equiv \rho+p_r+2p_t$, 
EC2$\equiv \rho+p_r$ and EC3$\equiv \rho+p_t$, for $\gamma=3$,
$\omega=-2/3$, $a=0.01$ and $m_c=0.1054688609\times 10^{-8}$. The horizons are: $r_h=7.071067814$; $r_{bh}=-0.2368329842\times 10^{-7}$, $r_c=99.99999998$ ($L_e=100$); $r_{bh}=-0.1657830889\times 10^{-7}$, $r_c=70$ ($L_e=70$); $r_{bh}=-0.1456327969\times 10^{-7}$, $r_c=61.49177126$ ($L_e=61.49177124$). {\bf Case F}}
\label{fig13}
\end{figure}

\begin{figure}
\vspace{.2in}
\centerline{\psfig{figure=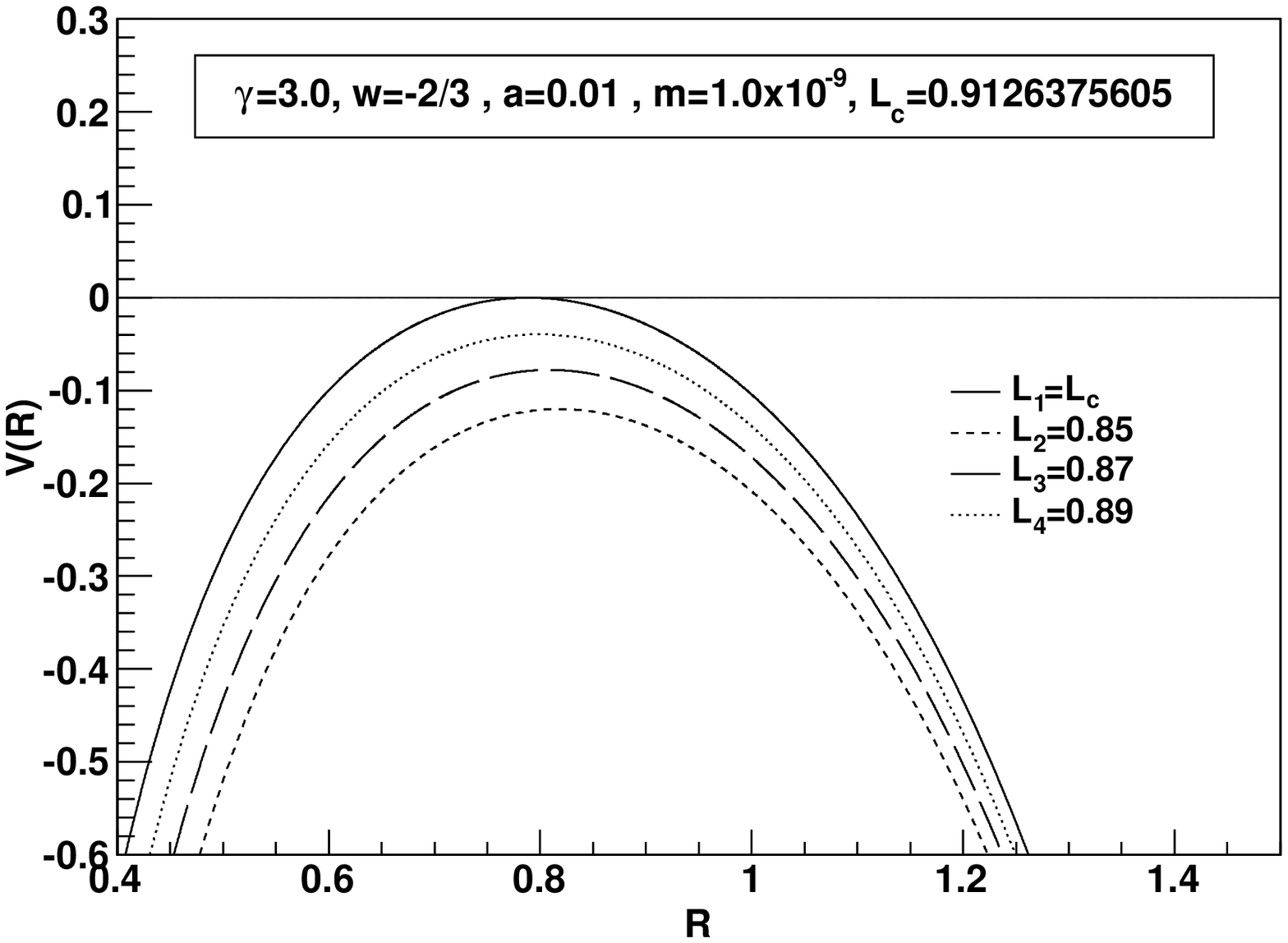,width=3.3truein,height=3.0truein}\hskip
.25in \psfig{figure=CondEnwm2div3a0v01.eps,width=3.3truein,height=3.0truein}
\hskip .5in} \caption{The potential $V(R)$ and the energy conditions EC1$\equiv \rho+p_r+2p_t$, 
EC2$\equiv \rho+p_r$ and EC3$\equiv \rho+p_t$, for $\gamma=3$,
$\omega=-2/3$, $a=0.01$ and $m_c=0.1054688609\times 10^{-8}$. The horizons are: $r_h=7.071067814$; $r_{bh}=0.1891503488\times 10^{-8}$, $r_c=0.9126375596$ ($L_e=0.9126375605$); $r_{bh}=0.1761682879\times 10^{-8}$, $r_c=0.8499999
992$ ($L_e=0.85$). {\bf Case F}}
\label{fig14}
\end{figure}

\begin{figure}
\vspace{.2in}
\centerline{\psfig{figure=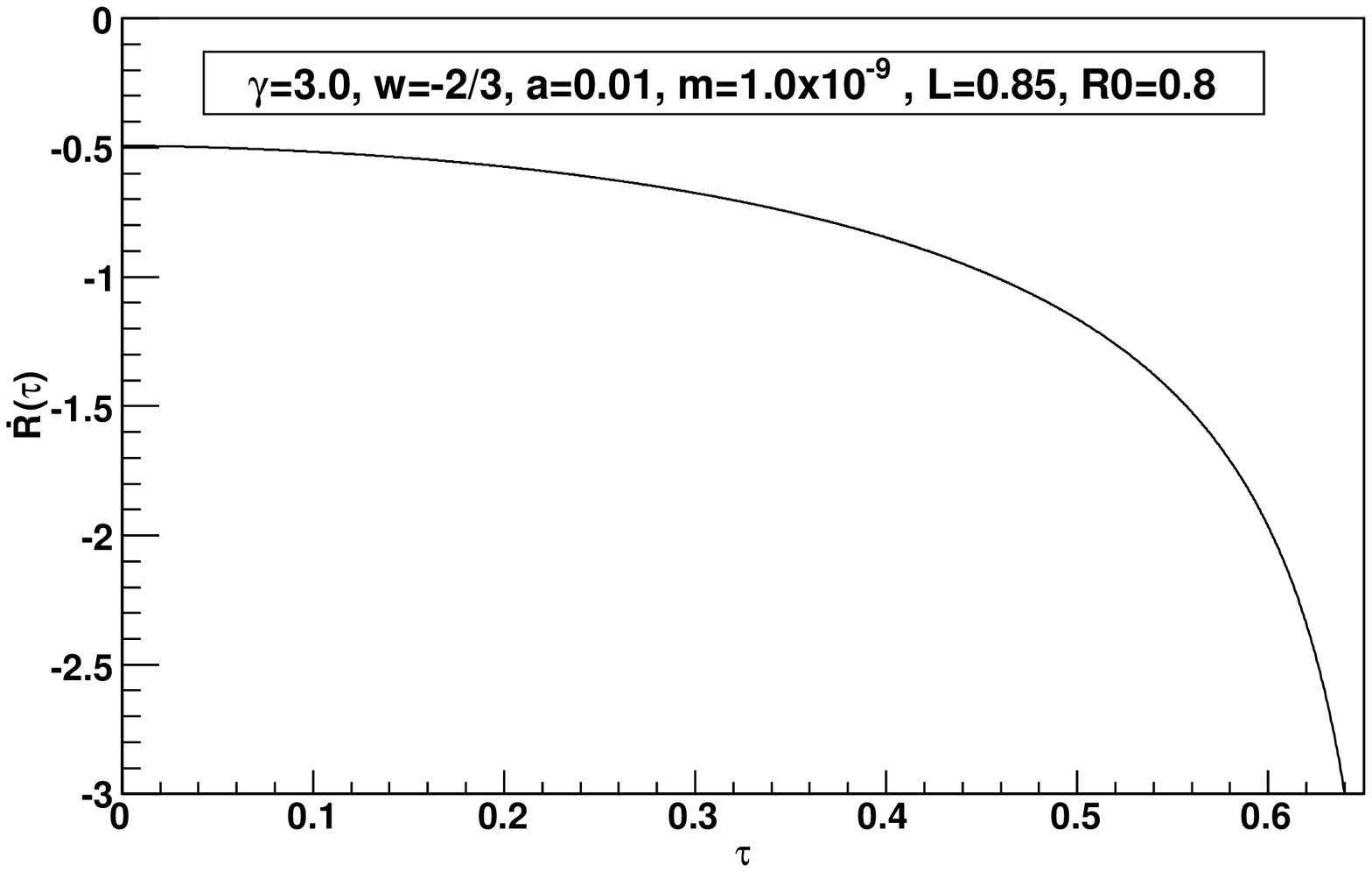,width=3.3truein,height=3.0truein}\hskip
.25in \psfig{figure=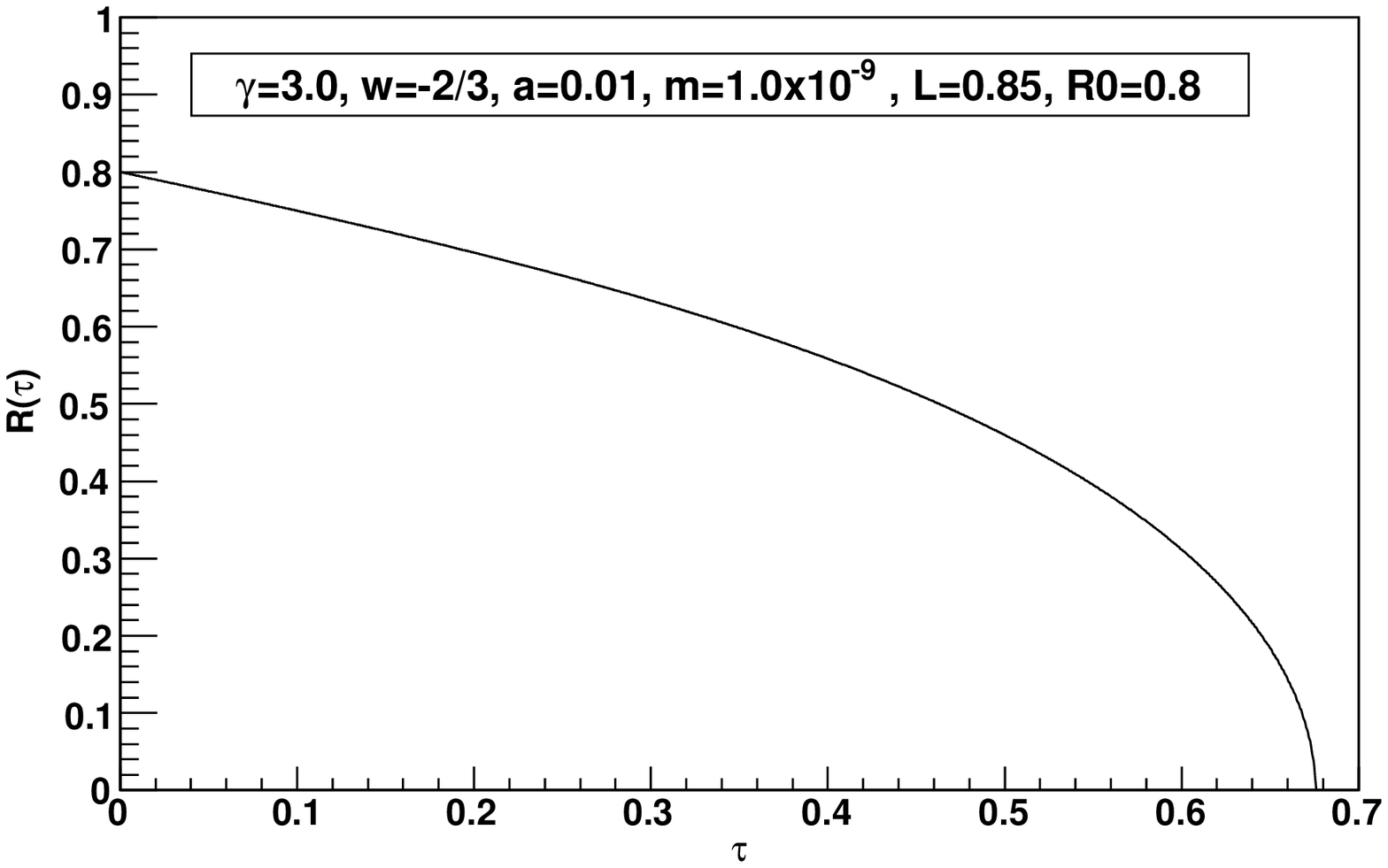,width=3.3truein,height=3.0truein}
\hskip .5in} \caption{These figures show the dynamical evolution of the collapse of a gravastar with the potential given by the figure \ref{fig14},
assuming $R_0=R(0)=0.9$, forming at the end of the evolution a black hole.} 
\label{fig15}
\end{figure}

\begin{figure}
\vspace{.2in}
\centerline{\psfig{figure=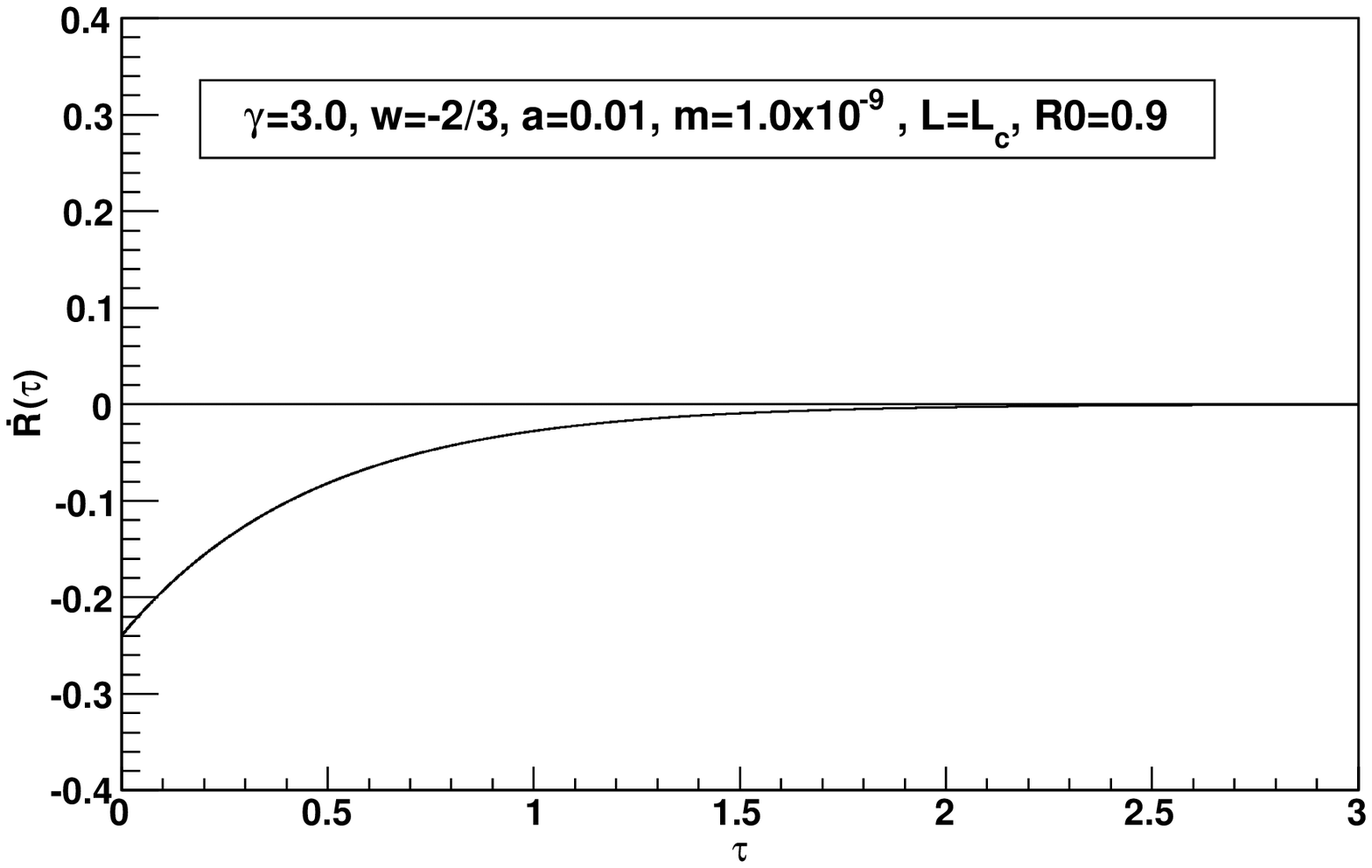,width=3.3truein,height=3.0truein}\hskip
.25in \psfig{figure=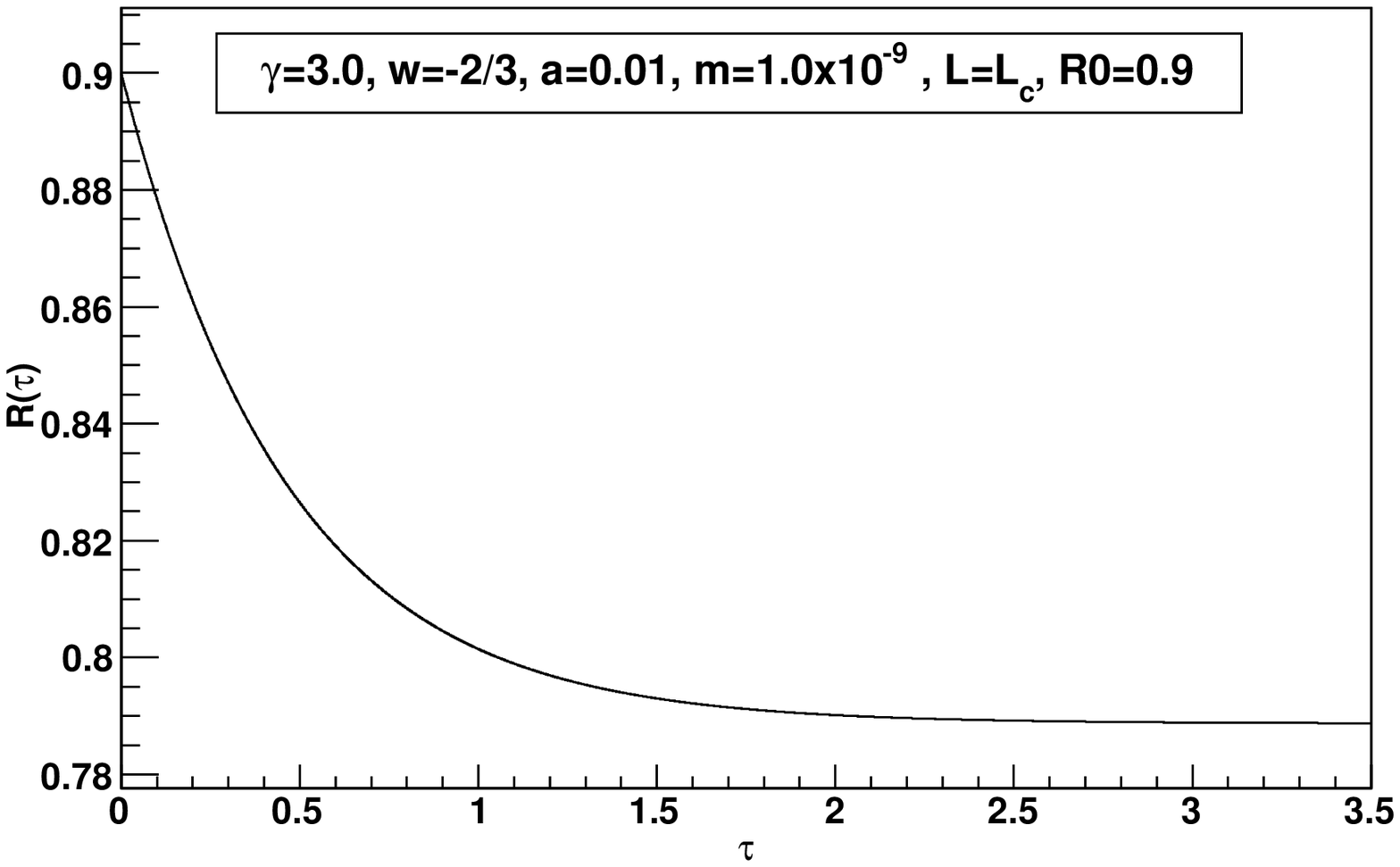,width=3.3truein,height=3.0truein}
\hskip .5in} \caption{These figures show the dynamical evolution of an unstable gravastar with the potential given by the figure \ref{fig14},
assuming $R_0=R(0)=0.9$.}
\label{fig16}
\end{figure}

\begin{figure}
\vspace{.2in}
\centerline{\psfig{figure=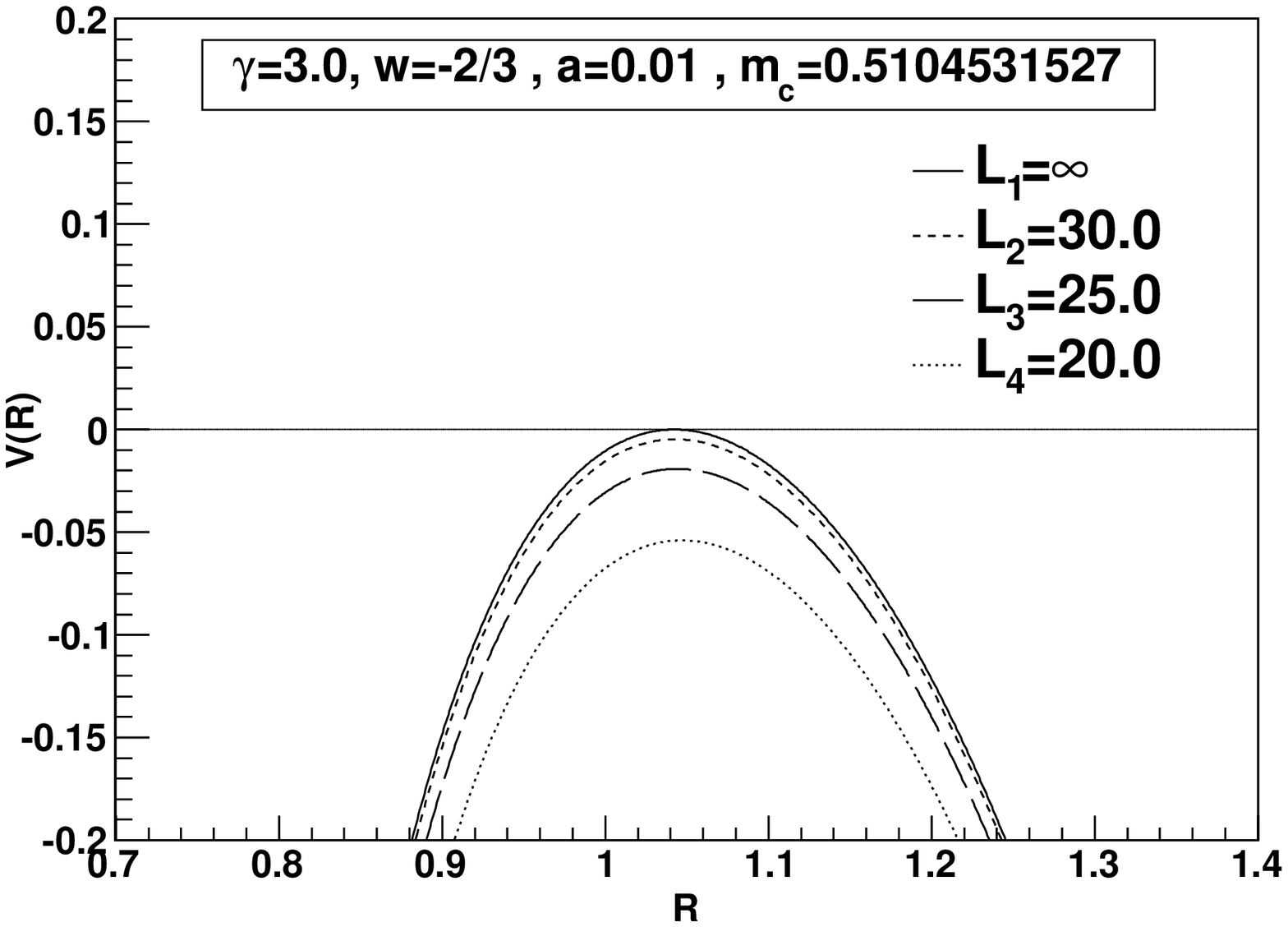,width=3.3truein,height=3.0truein}\hskip
.25in \psfig{figure=CondEnwm2div3a0v01.eps,width=3.3truein,height=3.0truein}
\hskip .5in} \caption{The potential $V(R)$ and the energy conditions EC1$\equiv \rho+p_r+2p_t$, 
EC2$\equiv \rho+p_r$ and EC3$\equiv \rho+p_t$, for $\gamma=3$,
$\omega=-2/3$, $a=0.01$ and $m_c=0.5104531527$. The horizons are: $r_h=7.071067814$; $r_{bh}=1.022092711$, $r_c=29.47589240$ ($L_e=30$); $r_{bh}=1.022617334$, $r_c=24.47300022$ ($L_e=25$); $r_{bh}=1.023587410$, $r_c=19.46855167$ ($L_e=20$). {\bf Case F}}
\label{fig17}
\end{figure}

\begin{figure}
\vspace{.2in}
\centerline{\psfig{figure=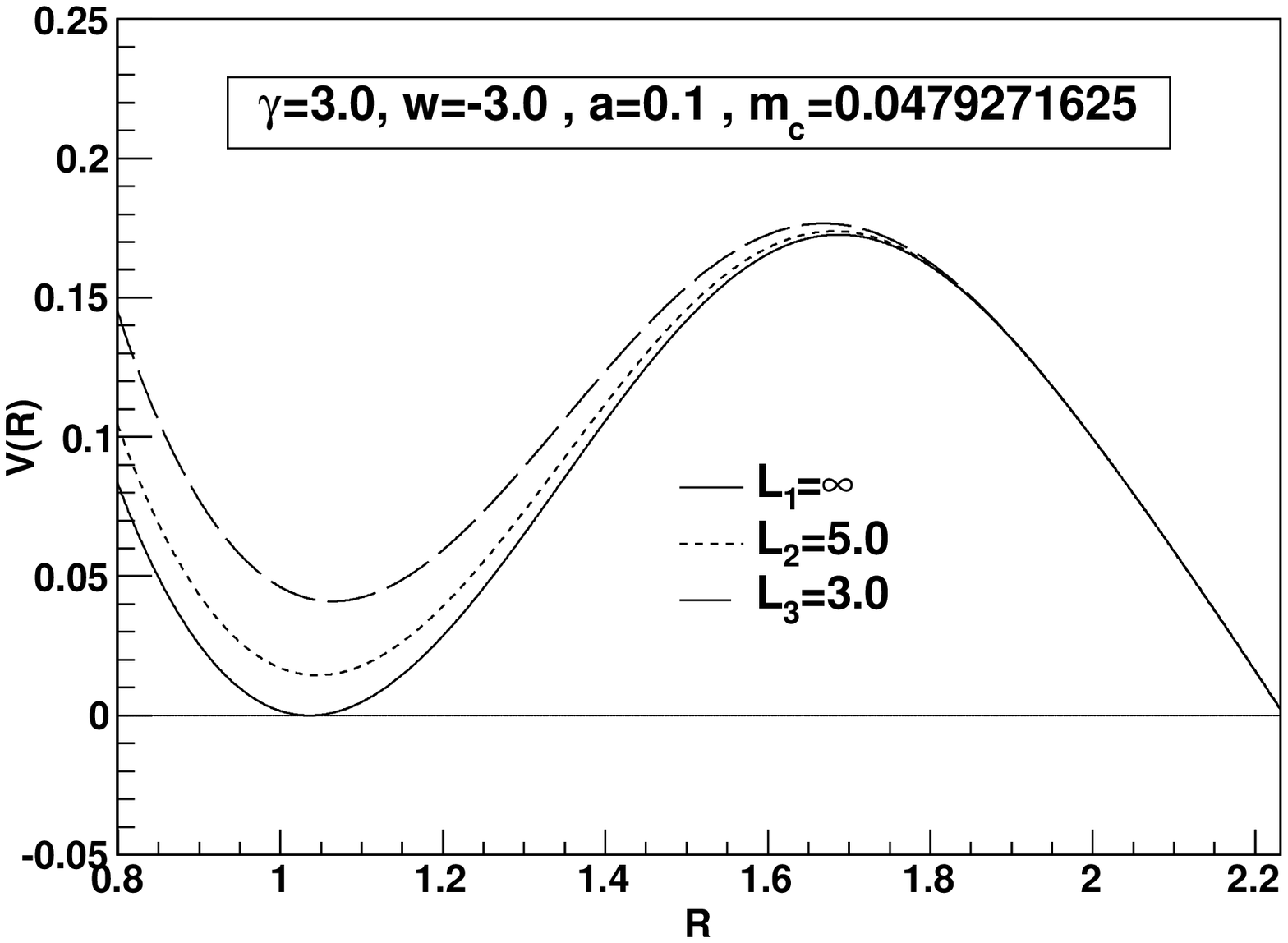,width=3.3truein,height=3.0truein}\hskip
.25in \psfig{figure=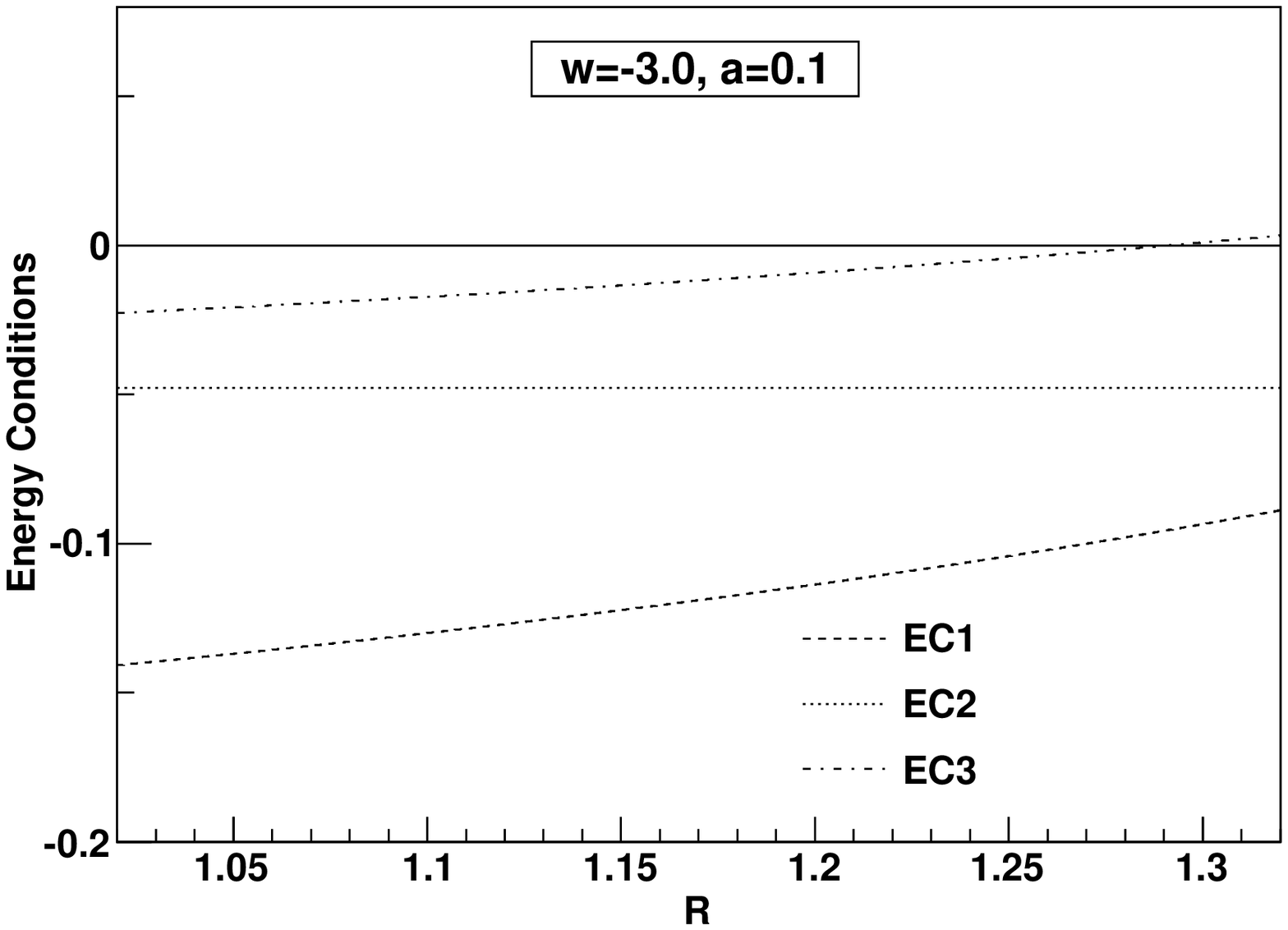,width=3.3truein,height=3.0truein}
\hskip .5in} \caption{The potential $V(R)$ and the energy conditions EC1$\equiv \rho+p_r+2p_t$, 
EC2$\equiv \rho+p_r$ and EC3$\equiv \rho+p_t$, for $\gamma=3$,
$\omega=-3$, $a=0.1$ and $m_c=0.0479271625$. The horizons are: $r_h=2.236067977$; $r_{bh}=0.09588959554$, $r_c=4.951365546$ ($L_e=5$); $r_{bh}=0.09595248530$, $r_c=2.950872678$ ($L_e=3$). {\bf Case I}}
\label{fig18}
\end{figure}

\begin{figure}
\vspace{.2in}
\centerline{\psfig{figure=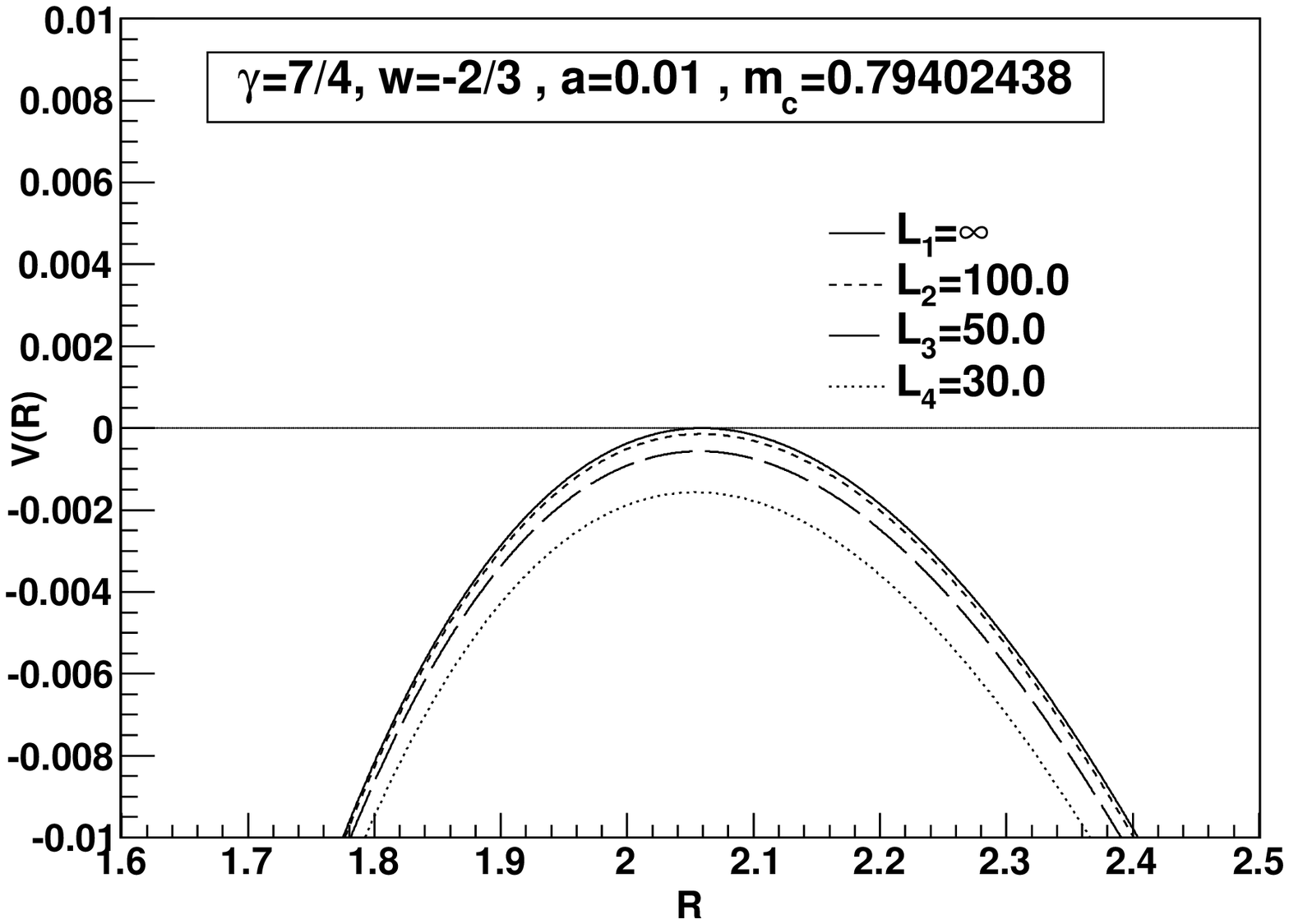,width=3.3truein,height=3.0truein}\hskip
.25in \psfig{figure=CondEnwm2div3a0v01.eps,width=3.3truein,height=3.0truein}
\hskip .5in} \caption{The potential $V(R)$ and the energy conditions EC1$\equiv
\rho+p_r+2p_t$,
EC2$\equiv \rho+p_r$ and EC3$\equiv \rho+p_t$, for $\gamma=7/4$,
$\omega=-2/3$, $b=0.01$ and $m_c=0.79402438$. The horizons are: $r_h=7.071067814
$; $r_{bh}=1.588449580$, $r_c=99.19631286$ ($L_e=100$); $r_{bh}=1.589655604$, $r
_c=49.18621608$ ($L_e=50$); $r_{bh}=1.592536495$, $r_c=29.17201284$ ($L_e=30$).
{\bf Case E}}
\label{fig20}
\end{figure}

\begin{figure}
\vspace{.2in}
\centerline{\psfig{figure=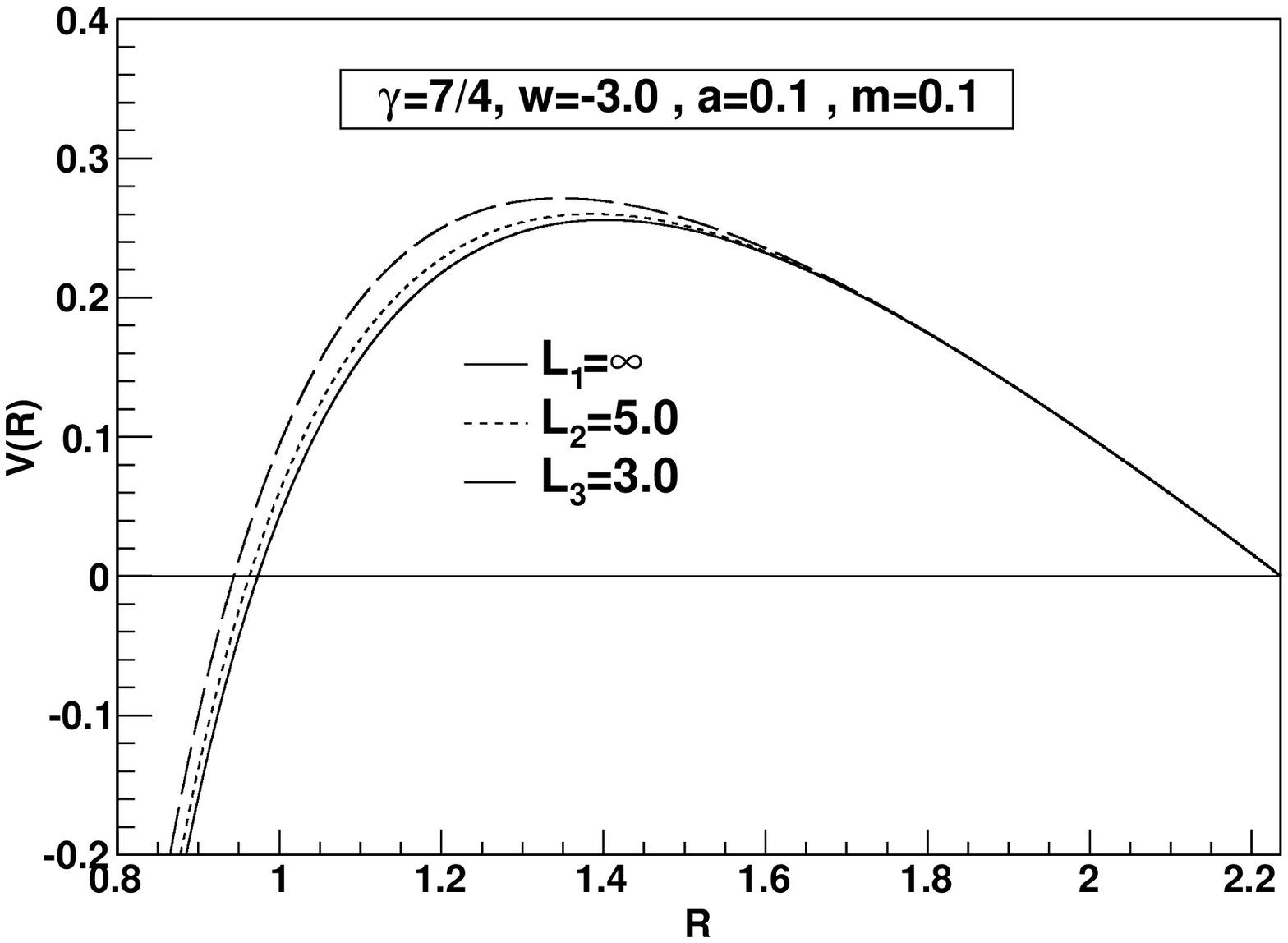,width=3.3truein,height=3.0truein}\hskip
.25in \psfig{figure=energycondwm3a0v1.eps,width=3.3truein,height=3.0truein}
\hskip .5in} \caption{The potential $V(R)$ and the energy conditions EC1$\equiv \rho+p_r+2p_t$, 
EC2$\equiv \rho+p_r$ and EC3$\equiv \rho+p_t$, for $\gamma=7/4$,
$\omega=-3$, $a=0.1$ and $m_c=0.1$. The horizons are: $r_h=7.071067814$; $r_{bh}=0.2003215448$, $r_c=4.896828668$ ($L_e=5$); $r_{bh}=0.2009009580$, $r_c=2.894500124$ ($L_e=3$). {\bf Case H}}
\label{fig21}
\end{figure}

\begin{acknowledgments}
The financial assistance from 
FAPERJ/UERJ (MFAdaS) are gratefully acknowledged. The
author (RC) acknowledges the financial support from FAPERJ (no.
E-26/171.754/2000, E-26/171.533/2002 and E-26/170.951/2006). 
The authors (RC and MFAdaS) also acknowledge the financial support from 
Conselho Nacional de Desenvolvimento Cient\'{\i}fico e Tecnol\'ogico - 
CNPq - Brazil.  The author (MFAdaS) also acknowledges the financial support
from Financiadora de Estudos e Projetos - FINEP - Brazil (Ref. 2399/03).
\end{acknowledgments}

\end{document}